\documentclass{article}
\usepackage{authblk}
\usepackage{amsmath, amssymb}
\usepackage{graphicx}
\usepackage{hyperref}
\usepackage{graphics}
\usepackage{rotating}
\usepackage{booktabs}
\usepackage{subcaption}
\usepackage{tikz-feynman}
\usetikzlibrary{feynman}

\begin{document}

\title{Static and dynamic properties of Triply Heavy Baryons}

\author{Kinjal Patel \thanks{kinjal1999patel@gmail.com}}

\author{Kaushal Thakkar\thanks{Corresponding Author: kaushal2physics@gmail.com}}

\affil{Department of Physics, Government College Daman, 396210, U. T. of Dadra \& Nagar Haveli and Daman \& Diu
           \and
           Veer Narmad South Gujarat University, Surat, India
}
 \maketitle

\begin{abstract}
\begin{sloppypar}
In this study, we investigate the ground-state masses, magnetic moments, transition magnetic moments, radiative decays, and heavy-to-heavy semileptonic decay rates, including their corresponding branching fractions of triply heavy baryons (THBs). The ground-state masses of the involved baryons are evaluated by numerically solving the six-dimensional hyperradial Schr\"{o}dinger equation within the hypercentral constituent quark model (hCQM), incorporating both hyper-Coulomb and linear confinement potentials along with spin-dependent interactions. The electromagnetic properties are calculated using the spin-flavour wave functions and the effective constituent quark masses of the baryon. The semileptonic $b \rightarrow  c$ decay widths are computed using the Isgur--Wise function within the heavy-quark spin symmetry, from which the corresponding branching ratios and lepton flavour universality ratios are also determined.
\end{sloppypar}
\end{abstract}
\begin{sloppypar}
\section{Introduction}\label{sec_i}
The study of hadrons containing heavy quarks provides an important testing ground for the dynamics of quantum chromodynamics (QCD) in both perturbative and non-perturbative regimes. In particular, heavy hadrons provide a unique opportunity to investigate quark confinement, heavy-quark symmetry, and the interplay between quark masses and strong interactions. Triply heavy baryons (THBs) composed of three heavy quarks (charm or bottom) represent a unique frontier in hadron physics. Although no such state has yet been observed experimentally, the discovery of the doubly charmed baryon $\Xi^{++}_{cc}$ by the LHCb Collaboration~\cite{Aaij2017,Aaij2018,Aaij2019,Aaij2020,Aaij2020a} significantly strengthened the prospects for observing triply heavy systems and renewed theoretical interest in their spectroscopy and decay properties. Triply heavy baryons are primarily governed by perturbative QCD due to the presence of three heavy quarks. Their large quark masses suppress the relativistic effects, making these systems comparatively simple to study while preserving important non-perturbative features, such as confinement and spin-dependent interactions. Consequently, triply heavy baryons provide an excellent testing ground for understanding QCD in the heavy-quark limit. Therefore, reliable theoretical predictions of their properties are essential to guide the experimental searches for these elusive states.

Despite the absence of direct experimental observations, theoretical estimates of the production rates of triply heavy baryons offer some guidance on the feasibility of future searches. The production of the triply charmed baryon in $e^+e^-$ collisions was estimated in Ref.~\cite{Baranov2004}. The production cross sections of triply heavy baryons at the LHC have been studied through the heavy-quark fragmentation in Refs.~\cite{Saleev1999,Nobary2005,Nobary2006,Nobary2006a}. The total and differential cross sections for the direct production of these baryons at the LHC were calculated in Refs.~\cite{Chen2011,Wu2012}, and the results indicate that the discovery of triply heavy baryons at the LHC is promising. The production of $\Omega_{ccc}$ in heavy-ion collisions was investigated in Ref.~\cite{He2015}. There, the production cross-section for $\Omega_{ccc}$ per binary collision in a central Pb+Pb collision at $\sqrt{s_{NN}}=2.76$ TeV reaches 9 nb, which is at least two orders of magnitude larger than that in a p+p collision at the same energy. This makes heavy-ion collisions at the LHC the most probable discovery channel, with their observation serving as a signature of quark-gluon plasma formation. In Ref.~\cite{Ghasemi2021}, the authors studied the influence of the spin orientation of heavy quarks in the production of triply heavy baryons through fragmentation in leading-order perturbative QCD. Ref.~\cite{Wu2023} reports that it is impossible to find triply heavy baryons at SuperKEKB, and that it is difficult to find $\Omega_{ccc}$ and $\Omega_{bbb}$ at the CEPC due to low event yields. While these production studies indicate that discovery may remain challenging in the near term, the theoretical properties of triply heavy baryons, such as their masses, decay widths, and internal structures, must be established before any experimental search.

Triply heavy baryons have been extensively studied using various theoretical frameworks, including lattice QCD~\cite{Brown2014,Meinel2010,Meinel2012,Padmanath2014,Dhindsa2025}, relativistic quark models~\cite{Faustov2022,Migura2006,Martynenko2008}, non-relativistic quark models~\cite{Pacheco2023,Patel2009,Vijande2015,Arenaza2024,Zhou2025,Cakir2026}, QCD sum rules~\cite{Aliev2014,Aliev2013,Najjar2024}, the bag model~\cite{Bernotas2009,Hasenfratz1980}, Faddeev equations~\cite{Radin2014,Guerrero2019,Qin2019,Yin2019}, the light-front quark model~\cite{Lu2024,Zhao2025,Wang2022}, and flavour SU(3) analysis~\cite{Huang2021}, and quark-diquark model~\cite{Bokade2026}. Most studies in the literature focus on spectroscopy, ground- and excited-state mass predictions, with comparatively fewer studies addressing semileptonic decays of triply heavy baryons. Existing studies on semileptonic decays have primarily focused on triply-to-doubly heavy baryon transitions rather than triply-to-triply heavy baryon transitions~\cite{Lu2024,Zhao2025,Wang2022,Najjar2025}. To our knowledge, the only predictions for $b \rightarrow  c$ semileptonic decays in triply-to-triply heavy baryon transitions are those of Ref.~\cite{Flynn2012}, obtained from heavy-quark spin symmetry combined with explicit non-relativistic wave function overlaps. Ref.~\cite{Wang2018} reported the lifetimes, as well as the semileptonic and nonleptonic weak decays of triply heavy baryons. In this study, we focus on the heavy-to-heavy $b \rightarrow  c$ semileptonic transitions of triply heavy baryons.

In this study, we extend our previous work~\cite{Patel2025} on the exclusive semileptonic decays of doubly heavy baryons. Here, we investigate the ground-state masses, magnetic moments, transition magnetic moments, and radiative and semileptonic decays of triply heavy baryons. This paper is organised as follows: In Section \ref{sec:1}, we calculate the ground-state masses of the triply heavy baryons in the hCQM. The magnetic moments and radiative decays are discussed in Section \ref{sec:2}. The Isgur--Wise function (IWF) and $b \rightarrow c$ semileptonic decay widths are computed in Section \ref{sec:3}. The results are presented and discussed in Section \ref{sec:4}.

\section{Masses of Triply Heavy Baryons}\label{sec:1}
We adopt the hypercentral constituent quark model (hCQM) to study the triply heavy baryons. We numerically solve the six-dimensional Schr\"{o}dinger equation to obtain the masses of the baryons using Wolfram Mathematica notebook \cite{Lucha1999}. In hCQM, Jacobi coordinates are essential for simplifying the three-body problem and providing a straightforward representation of inter-quark dynamics. Jacobi coordinates provide the relevant degrees of freedom for the relative motion of the three constituent quarks and are given as follows
\begin{equation}\label{eq:1}
\boldsymbol{\rho} = \frac{1}{\sqrt{2}}(\mathbf{r}_1 - \mathbf{r}_2),
\end{equation}
\begin{equation}\label{eq:2}
\boldsymbol{\lambda} = \frac{{m_1 \mathbf{r}_1 + m_2 \mathbf{r}_2 -  (m_1 + m_2) \mathbf{r}_3}}{{\sqrt{m_1^2 + m_2^2 + (m_1 + m_2)^2}}}.
\end{equation}
The reduced masses are given as
\begin{equation}\label{eq:3}
m_{\rho}=\frac{2 m_{1} m_{2}}{m_{1}+ m_{2}},
\end{equation}
\begin{equation}\label{eq:4}
m_{\lambda}=\frac{2 m_{3} (m_{1}^2 + m_{2}^2+m_1m_2)}{(m_1+m_2)(m_{1}+ m_{2}+ m_{3})},
\end{equation}
where $m_1$, $m_2$ and $m_3$ are the constituent quark masses. Unlike singly and doubly heavy baryons, all
three constituents entering Eqn.~(\ref{eq:3}) and (\ref{eq:4}) are heavy ($m_Q \gg \Lambda_{\rm QCD}$), so no light-quark mass scale enters the Jacobi construction for the triply heavy sector. The hyperspherical coordinates are given by the angles $\Omega_\rho=(\theta_\rho,\phi_\rho)$ and $\Omega_\lambda=(\theta_\lambda,\phi_\lambda)$. The hyperradius $x$ and hyperangle $\xi$ are defined as
\begin{equation}\label{eq:5}
x=\sqrt{\rho^2+\lambda^2}; \quad \xi=\arctan\left(\frac{\rho}{\lambda}\right).
\end{equation}
 Using hyperspherical coordinates, the kinetic energy operator $P_x^2/2m$ of the three-body system can be written as~\cite{Thakkar2020}
\begin{equation}\label{eq:6}
\frac{P_x^2}{2m} = -\frac{1}{2m} \left(\frac{\partial^2}{\partial x^2} + \frac{5}{x} \frac{\partial}{\partial x} - \frac{L^2(\Omega_\rho,\Omega_\lambda,\xi)}{x^2}\right),
\end{equation}
where $m= \frac{2m_\rho m_\lambda}{m_\rho+m_\lambda}$ denotes the reduced mass. $L^2(\Omega_\rho,\Omega_\lambda,\xi)$ is the quadratic Casimir operators of the six-dimensional rotational group $O(6)$ and its eigenfunctions are the hyperspherical harmonics $Y_{[\gamma]l_\rho l_\lambda}(\Omega_\rho,\Omega_\lambda,\xi)$ which satisfy the eigenvalue relation \cite{Patel2025a}
\begin{equation}\label{eq:7}
L^2Y_{[\gamma]l_{\rho}l_{\lambda}}(\Omega_{\rho},\Omega_{\lambda},\xi)=\gamma(\gamma+4)Y_{[\gamma]l_{\rho}l_{\lambda}}(\Omega_{\rho},\Omega_{\lambda},\xi),
\end{equation}
where $l_\rho$ and $l_\lambda$ are the angular momenta associated with the $\rho$ and $\lambda$ variables respectively. The model Hamiltonian for baryons can be expressed as
\begin{equation}\label{eq:8}
H = \frac{P_\rho^2}{2m_\rho}+\frac{P_\lambda^2}{2m_\lambda}+V(\rho,\lambda) = \frac{P_x^2}{2m} + V(x).
\end{equation}
Here, the potential $V(x)$ is not purely a two-body interaction but also includes three-body effects. The six-dimensional hyperradial Schr\"{o}dinger equation can be written as \cite{Patel2026}
\begin{eqnarray}\label{eq:9}
\left[ -\frac{1}{2m}\frac{d^2}{dx^2} + \frac{\frac{15}{4}+\gamma(\gamma + 4)}{2mx^2} + V(x) \right] \phi_{\gamma}(x) \nonumber \\
= E\phi_{\gamma}(x),
\end{eqnarray}
where $\phi_{\nu\gamma}(x) = x^{\frac{5}{2}}\psi_{\gamma}(x)$ is the hyperradial wave function and $\psi_{\gamma}(x)$ is the hypercentral wave function labelled by the grand angular quantum number $\gamma$ defined by the number of nodes $\nu$. The potential is assumed to depend only on the hyperradius and hence is a three-body potential because the hyperradius depends only on the coordinates of all three quarks. We consider the hypercentral potential $V(x)$ as the hyper-Coulomb (hC) plus linear potential, which is given as
\begin{equation}\label{eq:10}
V(x) = \frac{\tau}{x} + \beta x + V_0+ V_{spin},
\end{equation}\\
   where $\tau = -\frac{2}{3}\alpha_s$ is the hyper-Coulomb strength and the values of the potential parameters $\beta$ and $V_0$ are fixed to obtain the ground-state masses. $V_{spin}$ is the spin-dependent part that is perturbatively added and is given as~\cite{Garcilazo2007,Patel2025,majethiya2008a}
\begin{equation}\label{eq:11}
V_{spin}(x) = -\frac{A}{4} \alpha_s  \frac{e^{-x/x_0}}{x {x_0}^2} \sum_{i<j} \boldsymbol{\lambda}_i \cdot \boldsymbol{\lambda}_j \frac{\boldsymbol{\sigma}_i \cdot \boldsymbol{\sigma}_j}{6 m_i m_j}.
\end{equation}
Here, the parameter $A$ and the regularisation parameter $x_0$ are considered as the hyperfine parameters of the model. Parameter $x_0$ is treated as a hyperfine parameter related to gluon dynamics, independent of the masses of the interacting quarks, as proposed in Ref.~\cite{majethiya2008a}. There is no well-established procedure for evaluating $x_0$. $\boldsymbol{\lambda_{i,j}}$ are the SU(3) colour matrices, $\boldsymbol{\sigma_{i,j}}$ are the spin Pauli matrices, $m_{i,j}$ are the constituent masses of the two interacting quarks. Parameter $\alpha_s$ corresponds to the strong running coupling constant, which is given by
\begin{equation}\label{eq:12}
\alpha_s = \frac{\alpha_s(\mu_0)}{1+(\frac{33-2n_f}{12\pi})\alpha_s(\mu_0)\ln(\frac{m_1+m_2+m_3}{\mu_0})},
\end{equation}
     where $\alpha_s(\mu_0 = 1 GeV) \approx 0.6$ is considered in the present study. The masses of the ground-state $\Omega_{QQQ}$ baryons were calculated by summing the model quark masses and binding energy.
\begin{equation}\label{eq:13}
M_{QQQ} = m_1 + m_2 + m_3 + \langle H \rangle.
\end{equation}

\begin{table}
\centering
    \caption{\label{tab:Table1}Quark mass parameters and constants used in the calculations.}
    \begin{tabular}{cc}
    \hline\noalign{\smallskip}
    Parameter & Value\\
    \hline\noalign{\smallskip}
    ${m_{c}}$ & 1.55 GeV\\
    ${m_{b}}$ & 4.95 GeV\\
    $m_e$ & 0.000511 GeV\\
    $m_{\mu}$ & 0.1134 GeV\\
    $m_{\tau}$ & 1.776 GeV\\
    $\beta$ & 0.14 $GeV^2$\\
    ${V_0}$ & -0.90 GeV\\
    $x_0$ & 1.00 $GeV^{-1}$\\
    $\alpha_s(\mu_0 = 1 GeV)$ & 0.6\\
    \hline\noalign{\smallskip}
    \end{tabular}
\end{table}
\section{Electromagnetic Properties}\label{sec:2}
The electromagnetic properties of baryons are an important source of information regarding their internal structure. The magnetic moments of baryons are obtained in terms of the spin-flavour wave function of the constituent quarks as~\cite{majethiya2008}:
\begin{equation}\label{eq:13a}
\mu_B = \Sigma_i \langle \phi_{sf}|\mu_i|\phi_{sf}\rangle,
\end{equation}
where
\begin{equation}\label{eq:14}
\mu_i = \frac{e_i\sigma_i}{2m_i^{eff}}.
\end{equation}
where $i$ = c, b (and, for $\Omega_{ccb}$ and $\Omega_{bbc}$ states, both charm and bottom quarks appear among the constituents); $e_i$ and $\sigma_i$ represent the charge and spin of the constituting quarks of the baryonic state, respectively; $|\phi_{sf}\rangle$ represents the spin-flavour wave function of the respective baryonic state. Here, $m_i$, the mass of the $i^{th}$ quark in the three-body baryon, is taken as an effective mass of the constituting quarks, as their motions are governed by the three-body force described by the Hamiltonian in Eq.~(\ref{eq:6}). The baryon mass of the quarks may get modified due to their binding interactions with the other two quarks. We account for this bound-state effect by replacing the mass parameter $m_i$ in Eq.~(\ref{eq:14}) with an effective mass for the bound quarks,
$m_i^{eff}$, given as \cite{Thakkar2011}
\begin{equation}\label{eq:15}
m_i^{eff} = m_i \left( 1+\frac{\langle H \rangle}{\sum_i m_i}\right),
\end{equation}
such that $M_B = \sum_{i=1}^3 m_i^{eff}$, where $\langle H \rangle$ = $E$ + $\langle V(x) \rangle$. Since all three constituents of a triply heavy baryon are heavy quarks, the effective-mass correction in Eq.~(\ref{eq:15}) is applied uniformly to $m_c$ and/or $m_b$ depending on the flavour content of the state, in contrast to the doubly (and singly) heavy sectors where it is applied to a mixture of heavy and light quark masses. The calculated magnetic moments for THBs are listed and compared with other theoretical models in Table \ref{tab:Table3a}.\\

\subsection{Transition magnetic moment and radiative decay width}\label{sec:2a}
The transition magnetic moment for $\frac{3}{2}^+ \rightarrow \frac{1}{2}^+$ can be expressed as \cite{Thakkar2011}
\begin{equation}\label{eq:16}
\mu_{\frac{3}{2}^+ \rightarrow \frac{1}{2}^+} = \sum_i \left\langle
\phi_{sf}^{\frac{3}{2}^+} | \mu_i\sigma_i|
\phi_{sf}^{\frac{1}{2}^+} \right\rangle.
\end{equation}
$\langle \phi_{sf}^{\frac{3}{2}^+} |$ represents the spin-flavour wave function of the quark composition for the respective baryons with $J^P=\frac{3}{2}^+$, while $ |\phi_{sf}^{\frac{1}{2}^+}\rangle$ represents the spin-flavour wave function of the quark composition for the baryons with $J^P=\frac{1}{2}^+$. To compute the transition
magnetic moment ($\mu_{\frac{3}{2}^+ \rightarrow \frac{1}{2}^+}$), we take the geometric mean of the effective quark masses of the constituent quarks of the initial and final-state baryons,
\begin{equation}\label{eq:17}
m_i^{eff} = \sqrt{m_{i{B^*}}^{eff}m_{iB}^{eff}}.
\end{equation}
Here, $m_{i{B^*}}^{eff}$ and $m_{iB}^{eff}$ are the effective masses of the quarks constituting the baryonic states $B^*$ and $B$, respectively. Considering the geometric mean of the effective quark masses of the constituting quarks and the spin-flavour wave functions of the baryonic states, the transition magnetic moments are computed using Eq.~(\ref{eq:16}). The results are consistent with other theoretical predictions.

The radiative decay width can be expressed in terms of the radiative transition magnetic moment and photon momentum ($k$) as
\cite{Bernotas2013,Wagner2000}
\begin{equation}\label{eq:18}
 \Gamma = \frac{\alpha k^3}{M_P^2} \frac{2}{2J+1}\frac{M_B}{M_{B^*}}
 \mu^2(B^* \rightarrow B\gamma)
\end{equation}
where $\mu^2(B^* \rightarrow B\gamma)$ is the square of the transition magnetic moment, $\alpha=\frac{1}{137}$, and $M_P$ is the mass of the proton = 0.938~GeV. $J$ and $M_{B^*}$ are the total angular momentum and mass of the decaying baryon, respectively. $M_B$ is the final-state baryon mass. The photon momentum $k$ in the centre-of-mass system of the decaying baryon is given by
\begin{equation}\label{eq:19}
k = \frac{M^2_{B^*} - M^2_B}{2M_B^*}
\end{equation}
The obtained transition magnetic moments and radiative decay widths of the triply heavy baryons are listed in Table \ref{tab:Table4}.
\section{Isgur--Wise function and semileptonic decay}\label{sec:3}

For the semileptonic decay of triply heavy baryons, the general effective Hamiltonian for $b \rightarrow c$ transitions can be written as
\begin{equation}\label{eq:Heff_general}
\mathcal{H}_{eff} = \frac{G_F}{\sqrt{2}}V_{cb}\, \bar{c}\gamma_\mu(1-\gamma_5)b\, \bar{\ell}\gamma^\mu(1-\gamma_5)\nu_{\ell},
\end{equation}

where $G_F$ is the Fermi coupling constant, and $V_{cb}$ denotes the relevant Cabibbo--Kobayashi--Maskawa (CKM) matrix element.
\begin{equation}\label{eq:amp_general}
\begin{split}
M = \langle \Omega_f | \mathcal{H}_{eff} | \Omega_i \rangle \\[2pt]
= \frac{G_F}{\sqrt{2}}V_{cb}\, \bar{\ell}\,\gamma^\mu(1-\gamma_5)\nu_{\ell}\,
\langle \Omega_f | \bar{c}\gamma_\mu(1-\gamma_5)b | \Omega_i \rangle,
\end{split}
\end{equation}

where the initial $\Omega_i$ and final $\Omega_f$ triply heavy baryon states are related by the replacement of one constituent $b$ quark by a $c$ quark,

\begin{equation}
\Omega^{-*}_{bbb} \rightarrow  \Omega^{0(*)}_{bbc}, \qquad
\Omega^{0(*)}_{bbc} \rightarrow  \Omega^{+(*)}_{bcc}, \qquad
\Omega^{+(*)}_{bcc} \rightarrow  \Omega^{++*}_{ccc}.
\label{eq:correspondence}
\end{equation}

Near the zero-recoil point, the separate heavy-quark spin symmetries of the initial and final triply heavy baryon states strongly constrain the hadronic matrix element of the weak current, reducing it to a single scalar Isgur--Wise function for each transition~\cite{Flynn2012}. This allows the semileptonic decay width to be written directly in terms of the Isgur--Wise function, the leptonic tensor, and kinematic factors, without requiring a separate set of form factors. The total decay width for a semileptonic $b\rightarrow  c$ transition between triply heavy baryon states is given by~\cite{Flynn2012}

\begin{equation}\label{eq:decay_width}
\Gamma = \frac{|V_{cb}|^2 G_F^2}{8\pi^4}\,\frac{m'^2}{m}
\int \sqrt{w^2-1}\; L^{\mu\nu}(q)\,H_{\mu\nu}(v,k)\, dw,
\end{equation}
where $m$ and $m'$ are the masses of the initial and final baryons, $w = \upsilon\cdot\upsilon'$ is the velocity transfer, and $q = p-p'$. Here, $G_F = 1.16\times10^{-5} GeV^{-2}$ is the Fermi coupling constant, $|V_{cb}|= 0.041$ is the CKM matrix element.
The leptonic tensor is \cite{Flynn2012}
\begin{equation}\label{eq:leptonic_tensor}
L^{\alpha\beta}(q) = A(q^2)\,g^{\mu\nu} + B(q^2)\,\frac{q^\mu q^\nu}{q^2},
\end{equation}
with
\begin{equation}\label{eq:AB}
\begin{split}
A(q^2) = \frac{I(q^2)}{6}\left(2q^2 - m_\ell^2 - \frac{m_\ell^4}{q^2}\right), \\[2pt]
B(q^2) = \frac{I(q^2)}{3}\left(q^2 + m_\ell^2 - 2\frac{m_\ell^4}{q^2}\right),
\end{split}
\end{equation}
\begin{equation}\label{eq:Iq2}
I(q^2) = \pi^2\, q^2\,(q^2 - m_\ell^2).
\end{equation}
\begin{equation}\label{eq:hadronic_tensor}
\begin{split}
H^{\alpha\beta}(v,k) = \frac{1}{2J+1}\sum_{r,r'} \langle \Omega_f, v,k,r' | j^\alpha(0) | \Omega_i, v,r \rangle \\[2pt]
\times \langle \Omega_f, v,k,r' | j^\beta(0) | \Omega_i, v,r \rangle^*,
\end{split}
\end{equation}
with $J$ the spin of the initial baryon, and $r$, $r'$ the helicities of the
initial and final baryon states, respectively. The small residual momentum of the final baryon $k$ vanishes at the exact zero-recoil point.
Using the heavy-quark spin symmetry relations in Ref.~\cite{Flynn2012}, the contracted product $L^{\alpha\beta}H_{\alpha\beta}$ for each transition family reduces to a compact expression involving a single Isgur--Wise function $\eta(\omega)$. The leptonic--hadronic tensor product $L^{\alpha\beta}H_{\alpha\beta}$ relations for each transitions are as follows \cite{Flynn2012}

\begin{equation}
\begin{split}
L^{\alpha\beta}H_{\alpha\beta}\Big|_{\Omega^*_{bbb}\rightarrow \Omega_{bbc}}
\approx \frac{8}{3}\,\eta^2(w)\,m m'(1+w)\\[2pt]
\times\Bigg[-3A(q^2)
+ B(q^2)\Big(\frac{(v\cdot q)^2}{q^2} - 1\Big)\Bigg]
\end{split}
\label{eq:32}
\end{equation}

\begin{equation}
\begin{split}
L^{\alpha\beta}H_{\alpha\beta}\Big|_{\Omega^*_{bbb}\rightarrow \Omega^*_{bbc}}
\approx \frac{1}{3}\,\eta^2(w)\,m m'
\Bigg[-8A(q^2)\,w(1+2w^2)\\[2pt]
+ B(q^2)\Big(-w(12+8w^2)
- 2\frac{(v\cdot q)(v'\cdot q)}{q^2}(20+8w^2)\Big)\Bigg]
\end{split}
\label{eq:LH_bbb_bbc}
\end{equation}

\begin{equation}
\begin{split}
L^{\alpha\beta}H_{\alpha\beta}\Big|_{\Omega_{bbc}\rightarrow \Omega_{bcc}}
\approx \frac{4}{9}\,\eta^2(w)\,m m'
\Bigg[-A(q^2)(34w+32)\\[2pt]
+ B(q^2)\Big[17\Big(2\frac{(v\cdot q)(v'\cdot q)}{q^2} - w\Big) - 8\Big]\Bigg]
\end{split}
\label{eq:28}
\end{equation}

\begin{equation}
\begin{split}
L^{\alpha\beta}H_{\alpha\beta}\Big|_{\Omega_{bbc}\rightarrow \Omega^*_{bcc}}
\approx \frac{16}{9}\,\eta^2(w)\,m m'(1+w)\\[2pt]
\times\Bigg[-3A(q^2)
+ B(q^2)\Big(\frac{(v'\cdot q)^2}{q^2} - 1\Big)\Bigg]
\end{split}
\label{eq:30}
\end{equation}

\begin{equation}
\begin{split}
L^{\alpha\beta}H_{\alpha\beta}\Big|_{\Omega^*_{bbc}\rightarrow \Omega_{bcc}}
\approx \frac{8}{9}\,\eta^2(w)\,m m'(1+w)\\[2pt]
\times\Bigg[-3A(q^2)
+ B(q^2)\Big(\frac{(v\cdot q)^2}{q^2} - 1\Big)\Bigg]
\end{split}
\label{eq:29}
\end{equation}

\begin{equation}
\begin{split}
L^{\alpha\beta}H_{\alpha\beta}\Big|_{\Omega^*_{bbc}\rightarrow \Omega^*_{bcc}}
\approx \frac{4}{9}\,\eta^2(w)\,m m'
\Bigg[-8A(q^2)\,w(1+2w^2)\\[2pt]
+ B(q^2)\Big(-w(12+8w^2)
+ 2\frac{(v\cdot q)(v'\cdot q)}{q^2}(20+8w^2)\Big)\Bigg]
\end{split}
\label{eq:31}
\end{equation}

\begin{equation}
\begin{split}
L^{\alpha\beta}H_{\alpha\beta}\Big|_{\Omega_{bcc}\rightarrow \Omega^*_{ccc}}
\approx \frac{16}{3}\,\eta^2(w)\,m m'(1+w)\\[2pt]
\times\Bigg[-3A(q^2)
+ B(q^2)\Big(\frac{(v'\cdot q)^2}{q^2} - 1\Big)\Bigg]
\end{split}
\label{eq:26}
\end{equation}

\begin{equation}
\begin{split}
L^{\alpha\beta}H_{\alpha\beta}\Big|_{\Omega^*_{bcc}\rightarrow \Omega^*_{ccc}}
\approx \frac{1}{3}\,\eta^2(w)\,m m'
\Bigg[-8A(q^2)\,w(1+2w^2)\\[2pt]
+ B(q^2)\Big(-w(12+8w^2)
+ 2\frac{(v\cdot q)(v'\cdot q)}{q^2}(20+8w^2)\Big)\Bigg]
\end{split}
\label{eq:27}
\end{equation}

We follow the phenomenological approach of Faessler et al.~\cite{Faessler2009}, originally developed for doubly heavy baryon transitions, and represent each Isgur--Wise function by an exponential form,

\begin{equation}\label{eq:iwf_ansatz}
\eta(w) = \exp\left(-3(w - 1)\frac{m_{Q}^2}{\Lambda_B^2}\right)
\end{equation}
 Here $\Lambda_B$ is a phenomenological size parameter that varies in the range $3 \le \Lambda_B \le 7~\text{GeV}$~\cite{Faessler2006}, parametrising the distribution of quarks inside a given baryon. This ansatz satisfies the heavy-quark spin symmetry normalisation $\eta(1)=1$ required at zero recoil.

\section{Results and Discussion}\label{sec:4}

The baryon masses were obtained by numerically solving the six-dimensional Schr\"{o}dinger equation within the hypercentral constituent quark model (hCQM), employing a potential composed of a hyper-Coulomb like term and a linear term. A spin-dependent interaction is included perturbatively to reproduce the observed mass splittings among baryon states. We have calculated the ground-state masses of all the triply heavy baryons using the parameters listed in Table \ref{tab:Table1}. We set the same parameters for all the THBs using which we evaluated various properties of THBs. The calculated masses of the ground-state THBs are presented in Table \ref{tab:Table2}. Table \ref{tab:Table2} summarises the results, compared with the predictions from lattice QCD~\cite{Brown2014}, the variational approach~\cite{Flynn2012}, and the relativistic quark
model~\cite{Faustov2022}. The calculated masses agree reasonably well with the other theoretical predictions for $\Omega^*_{bcc}$, $\Omega_{bcc}$ and $\Omega^*_{ccc}$ baryons. However, the largest deviation from lattice QCD is observed for $\Omega^{*}_{bbb}$, whose predicted mass of $14.852$~GeV lies about $3.4\%$ above the LQCD value of $14.366$~GeV~\cite{Brown2014}.

The magnetic moments of all the triply heavy baryons are computed using the spin-flavour wave functions and effective constituent quark masses of the corresponding baryon. The calculated magnetic moments are compared with other theoretical predictions in Table \ref{tab:Table3a}. Since the individual quark magnetic moment (see Eq. \ref{eq:14}) scales inversely with the effective quark mass, heavier constituents contribute less to the total baryon magnetic moment, whereas the sign of $\mu_i$ is fixed entirely by the quark charge $e_i$. As the mass of bottom quark $m_b$ is much greater than the mass of charm quark $m_c$ in the present parameter set, the charm quark dominates the magnetic moment whenever it is present, and this mass hierarchy is directly responsible for the systematic pattern observed in Table \ref{tab:Table3a}. The magnitude of the magnetic moment increases monotonically from $\Omega^*_{bbb}$ ($-0.189$) to $\Omega^*_{ccc}$ (1.166) as the charm content of the baryon increases, while the sign tracks the dominant heavy-quark charge, negative for bottom-rich states and positive for charm-rich states. The transition magnetic moments and the corresponding radiative $M1$ decay widths for the $\Omega^*_{bbc}\rightarrow\Omega_{bbc}\gamma$ and $\Omega^*_{bcc}\rightarrow\Omega_{bcc}\gamma$ transitions are listed in Table \ref{tab:Table4} and compared with the effective quark mass scheme~\cite{Hazra2021} and bag model~\cite{Simonis2018} predictions. Our predicted transition magnetic moments are in reasonable agreement with both references for the $\Omega^*_{bcc}\rightarrow\Omega_{bcc}\gamma$ transition, whereas for $\Omega^*_{bbc}\rightarrow\Omega_{bbc}\gamma$ our value ($-0.439$) is somewhat larger in magnitude than that predicted by \cite{Simonis2018}($-0.352$). Since the radiative decay widths follow the same trend as the transition magnetic moments, our predicted width for $\Omega^*_{bbc}\rightarrow\Omega_{bbc}\gamma$ (0.007~keV) is lower than both the predictions, while our width for $\Omega^*_{bcc}\rightarrow\Omega_{bcc}\gamma$ (0.0195~keV) agrees with that of Ref.~\cite{Hazra2021}.  The discrepancies in the radiative decay widths can also be attributed to the photon momentum $k$, as given by Eq. (\ref{eq:19}), which is sensitive to the masses of the baryons. Since $\Gamma \propto k^3$, even a modest difference in the predicted mass splitting between the initial and final states across different models can translate into a sizeable difference in the radiative decay width.

The exclusive semileptonic decay widths for the $b\rightarrow c$ transitions of the triply heavy baryon family are listed and compared in Table \ref{tab:Table5}. Given the near-complete absence of independent theoretical predictions for these decays and the current lack of experimental data on triply heavy baryons, this work tests whether the exponential Isgur--Wise function ansatz of Faessler et al.~\cite{Faessler2009}, previously applied by us to the doubly heavy baryon sector with a size parameter $\Lambda_B$ fitted to that sector~\cite{Patel2025}, extends consistently to the triply heavy sector. Here, the heavy-quark spin symmetry reduction of the hadronic matrix elements proceeds through the doubly heavy diquark subsystem that remains a spectator in each $b\rightarrow c$ transition, following the formalism of Ref.~\cite{Flynn2012}. Since the internal structure and effective size of a triply heavy baryon differ substantially from those of a doubly heavy baryon, owing to the additional heavy quark replacing what was previously a light or strange spectator, the $\Lambda_B$ value used in Eq.~(\ref{eq:iwf_ansatz}) cannot be carried over directly from the doubly heavy analysis. We therefore adopt a common value $\Lambda_B = 3$~GeV across all three transition families; a smaller $\Lambda_B$ yields a smaller decay width, and vice versa. The flavour dependence enters through $m_Q$, for the $\Omega_{bbb}\rightarrow\Omega_{bbc}$ transition, $m_Q = 3m_b$, since the initial baryon is composed entirely of bottom quarks; for the $\Omega_{bbc}\rightarrow\Omega_{bcc}$ transition, which connects two mixed-flavour baryons, $m_Q = \frac{3}{2}(m_b+m_c)$; and for the $\Omega_{bcc}\rightarrow\Omega_{ccc}$ transition, $m_Q = 3m_c$. With this refinement, the resulting semileptonic decay widths are systematically larger than those obtained in the doubly heavy sector~\cite{Patel2025}, a difference driven by both the larger phase space and velocity-transfer range available in these heavier systems and the flavour-dependent scaling of $m_Q$ in the Isgur--Wise function $\eta(w)$.

\begin{figure*}
\centering
\begin{subfigure}{0.45\textwidth}
\includegraphics[scale=0.3]{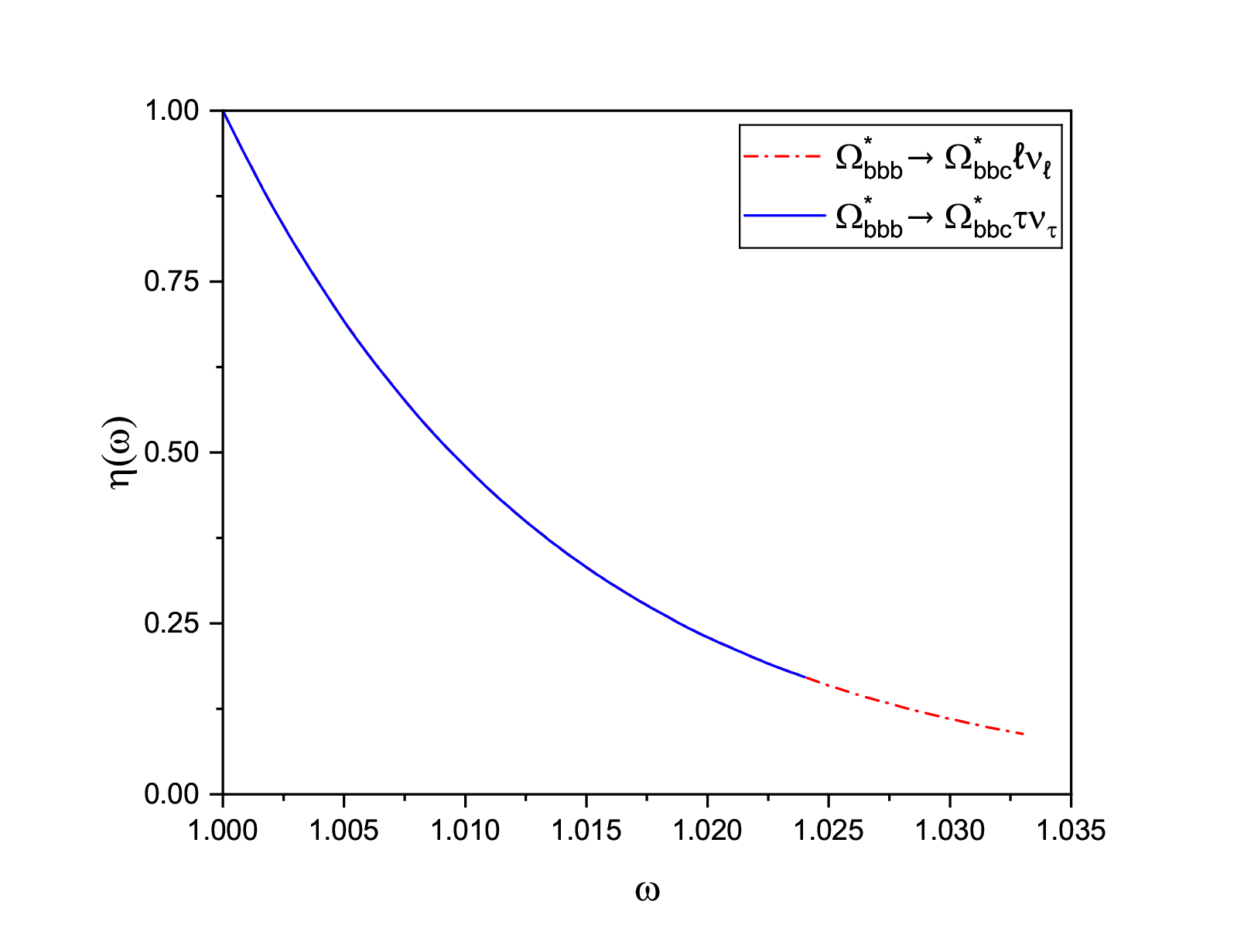}
\caption{$\Omega^*_{bbb} \rightarrow \Omega_{bbc}\,\ell\bar{\nu}$ with ($\ell=e,\mu$)}
\label{fig:1a}
\end{subfigure}
\hfill
\begin{subfigure}{0.45\textwidth}
\includegraphics[scale=0.3]{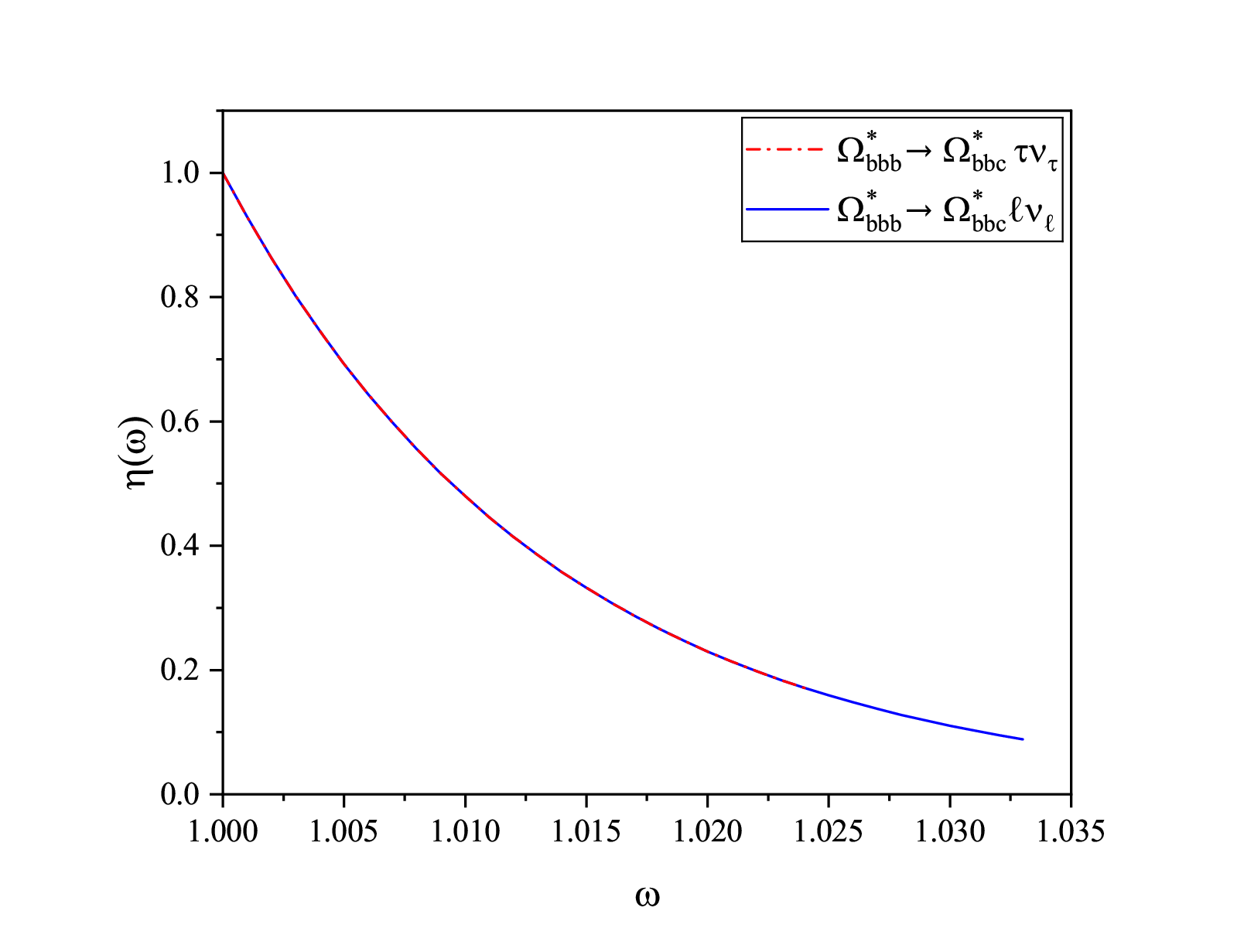}
\caption{$\Omega^*_{bbb} \rightarrow \Omega^*_{bbc}\,\ell\bar{\nu}$ with ($\ell=e,\mu$)}
\label{fig:1b}
\end{subfigure}
\caption{The Isgur--Wise function $\eta(\omega)$ for $\Omega^*_{bbb} \rightarrow \Omega^{(*)}_{bbc}\,\ell\bar{\nu}$ transitions}
\label{fig:1}
\end{figure*}

\begin{figure*}
\centering
\begin{subfigure}{0.45\textwidth}
\includegraphics[scale=0.3]{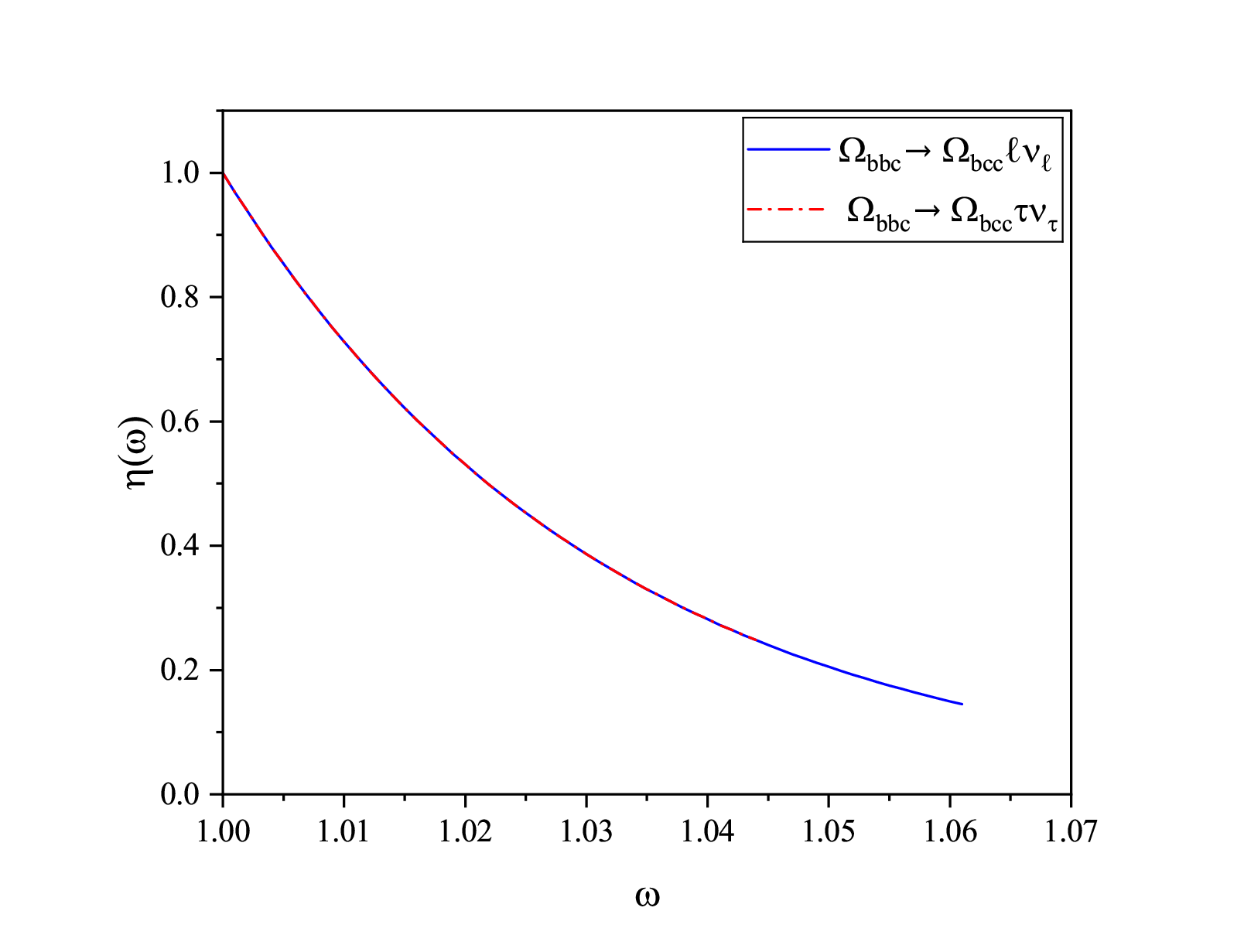}
\caption{$\Omega_{bbc} \rightarrow \Omega_{bcc}\,\ell\bar{\nu}$ with ($\ell=e,\mu$)}
\label{fig:2a}
\end{subfigure}
\hfill
\begin{subfigure}{0.45\textwidth}
\includegraphics[scale=0.3]{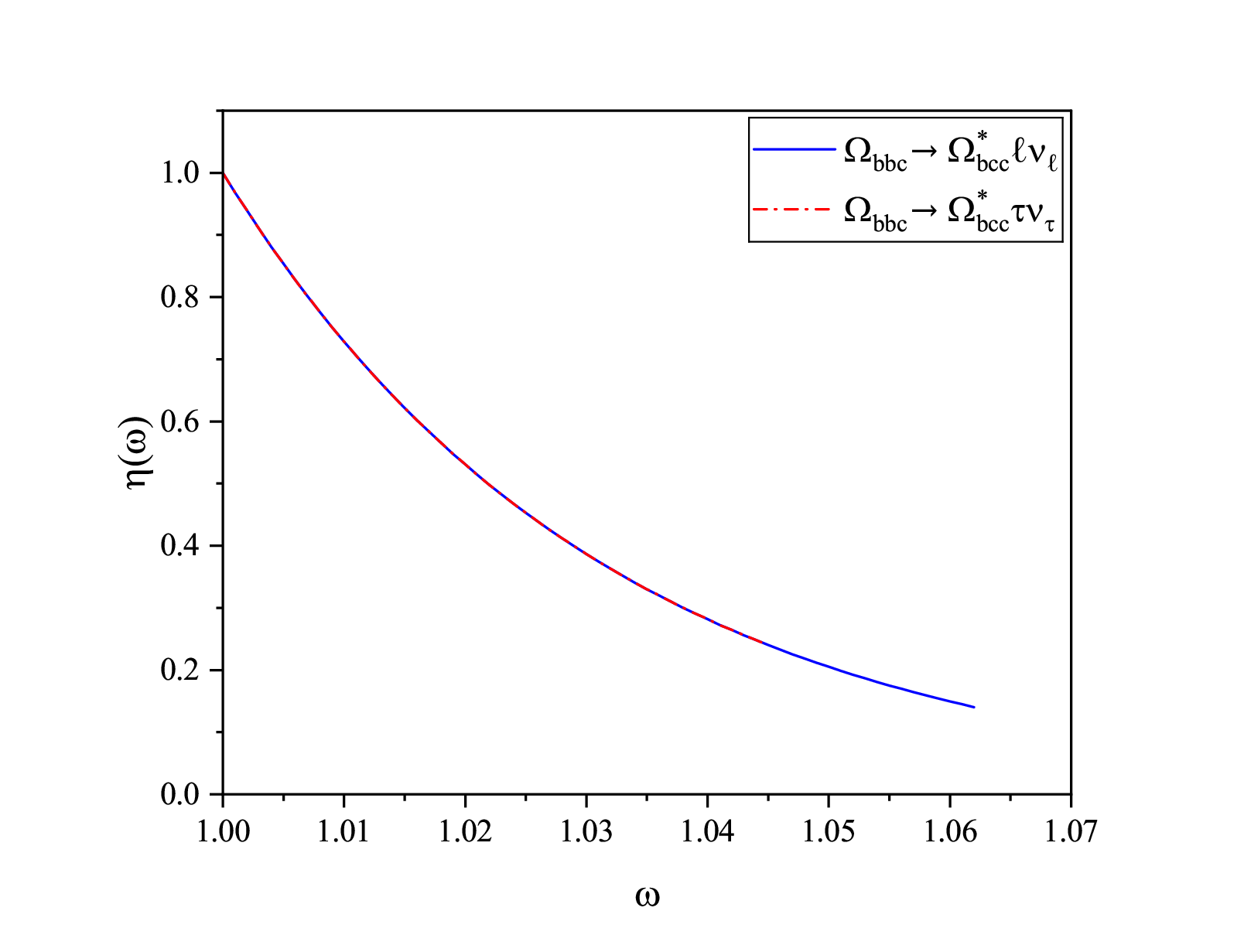}
\caption{$\Omega_{bbc} \rightarrow \Omega^*_{bcc}\,\ell\bar{\nu}$ with ($\ell=e,\mu$)}
\label{fig:2b}
\end{subfigure}
\caption{The Isgur--Wise function $\eta(\omega)$ for $\Omega_{bbc} \rightarrow \Omega^{(*)}_{bcc}\,\ell\bar{\nu}$ transitions}
\label{fig:2}
\end{figure*}

\begin{figure*}
\centering
\begin{subfigure}{0.45\textwidth}
\includegraphics[scale=0.3]{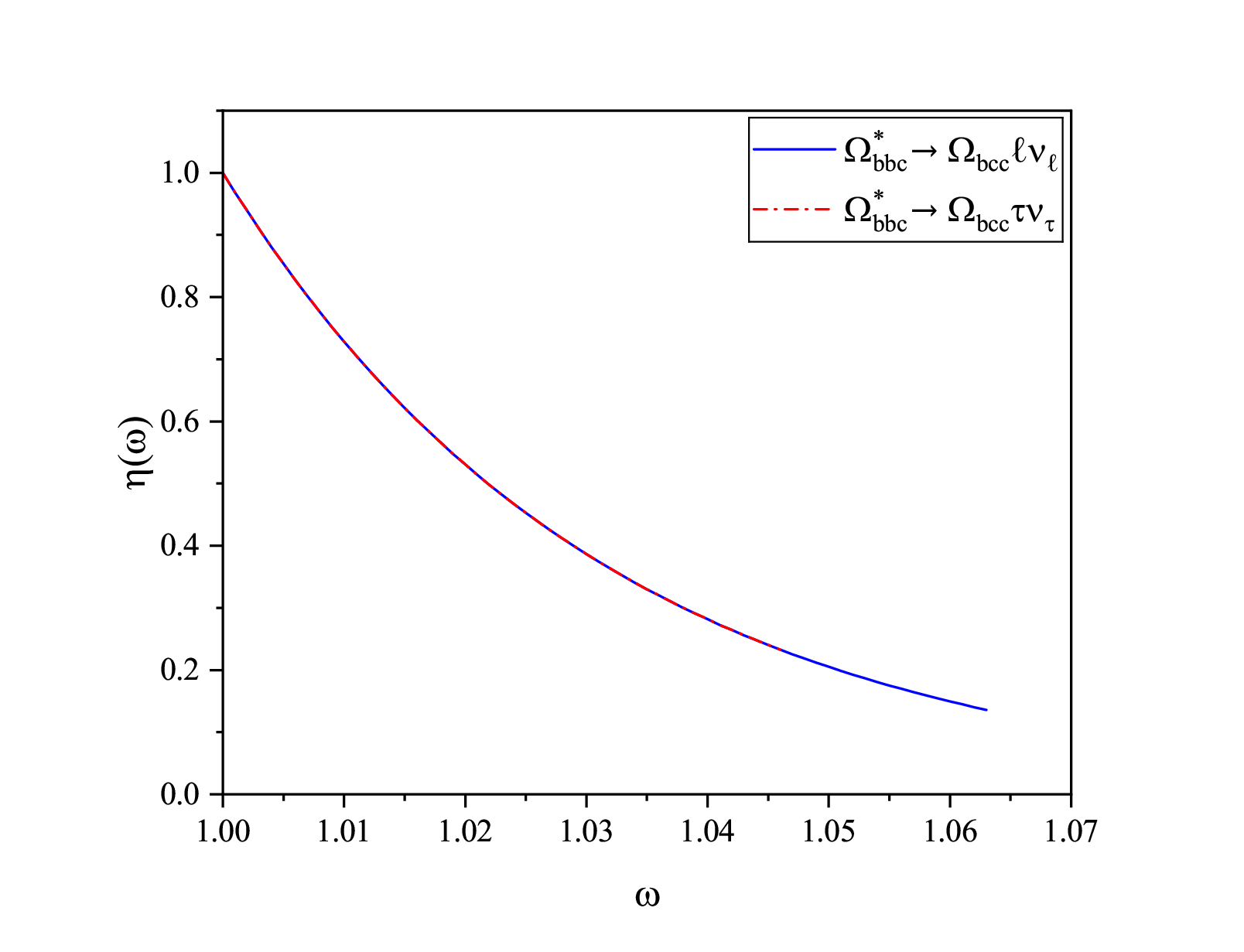}
\caption{$\Omega^*_{bbc} \rightarrow \Omega_{bcc}\,\ell\bar{\nu}$ with ($\ell=e,\mu$)}
\label{fig:3a}
\end{subfigure}
\hfill
\begin{subfigure}{0.45\textwidth}
\includegraphics[scale=0.3]{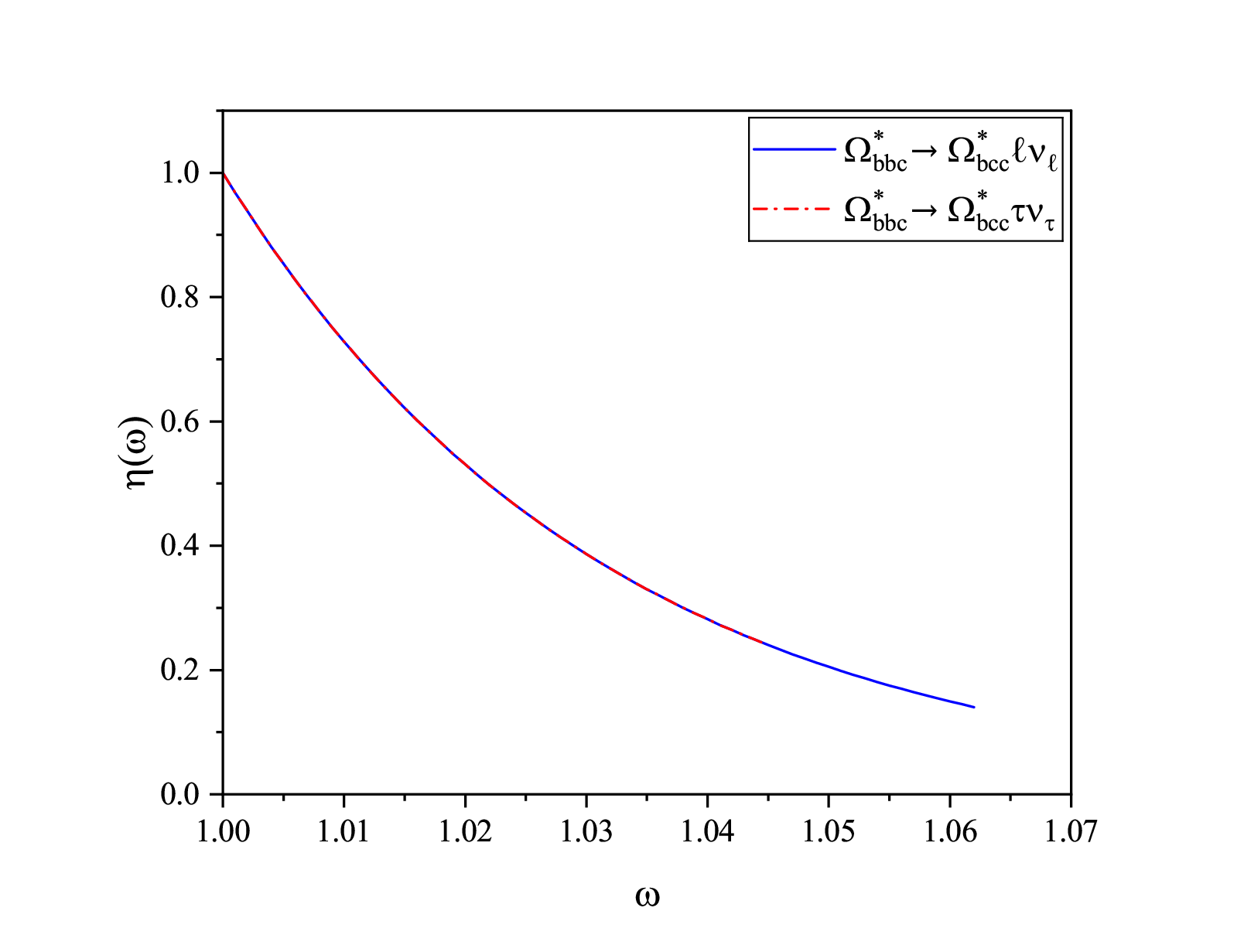}
\caption{$\Omega^*_{bbc} \rightarrow \Omega^*_{bcc}\,\ell\bar{\nu}$ with ($\ell=e,\mu$)}
\label{fig:3b}
\end{subfigure}
\caption{The Isgur--Wise function $\eta(\omega)$ for $\Omega^*_{bbc} \rightarrow \Omega^{(*)}_{bcc}\,\ell\bar{\nu}$ transitions}
\label{fig:3}
\end{figure*}

\begin{figure*}
\centering
\begin{subfigure}{0.45\textwidth}
\includegraphics[scale=0.3]{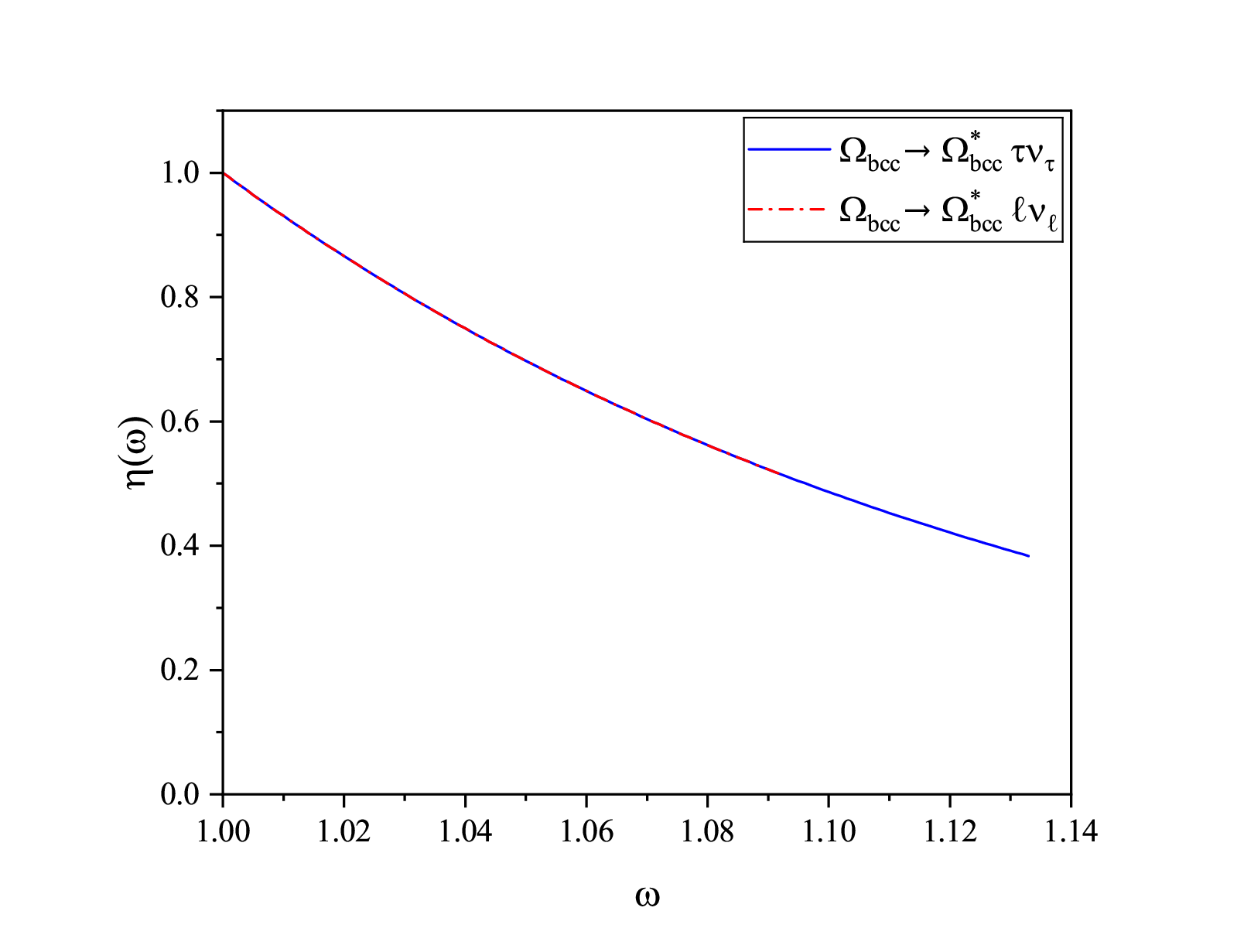}
\caption{$\Omega_{bcc} \rightarrow \Omega^*_{ccc}\,\ell\bar{\nu}$ with ($\ell=e,\mu$)}
\label{fig:4a}
\end{subfigure}
\hfill
\begin{subfigure}{0.45\textwidth}
\includegraphics[scale=0.3]{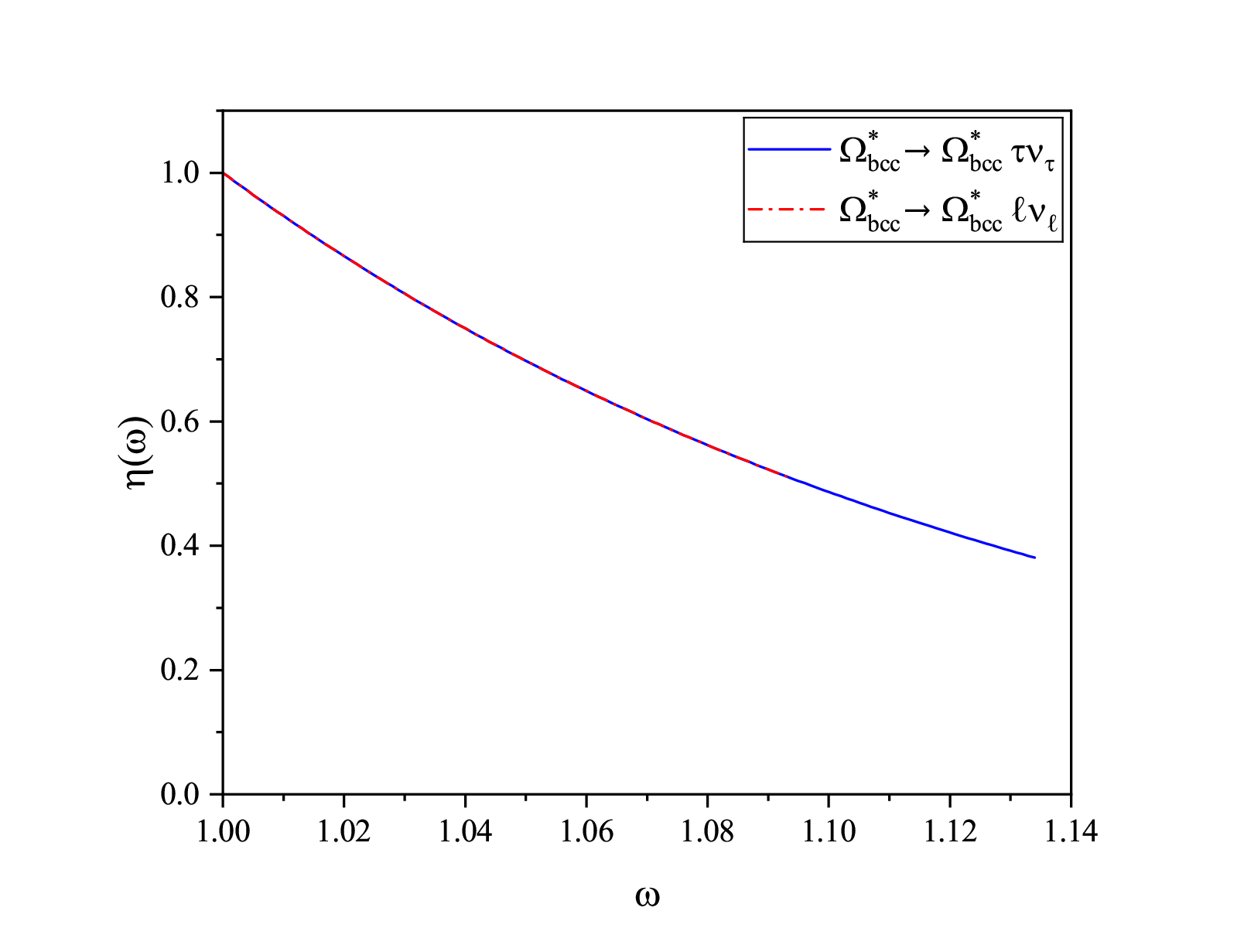}
\caption{$\Omega^*_{bcc} \rightarrow \Omega^*_{ccc}\,\ell\bar{\nu}$ with ($\ell=e,\mu$)}
\label{fig:4b}
\end{subfigure}
\caption{The Isgur--Wise
 function $\eta(\omega)$ for $\Omega^{(*)}_{bcc} \rightarrow \Omega^*_{ccc}\,\ell\bar{\nu}$ transitions}
\label{fig:4}
\end{figure*}

As $m_Q$ decreases from $3m_b = 14.85$~GeV ($\Omega_{bbb}^{*}\rightarrow \Omega_{bbc}$, Fig.~\ref{fig:1}) through $\frac{3}{2}(m_b+m_c) = 9.75$~GeV ($\Omega_{bbc}\rightarrow \Omega_{bcc}$, Figs.~\ref{fig:2} and \ref{fig:3}) to $3m_c = 4.65$~GeV ($\Omega_{bcc}\rightarrow \Omega_{ccc}$, Fig.~\ref{fig:4}), the kinematically accessible $\omega$ range widens substantially, from $1.000$--$1.035$ for the $bbb\rightarrow  bbc$ family to $1.00$--$1.14$ for the $bcc\rightarrow  ccc$ family, and the Isgur--Wise function falls off more gradually in absolute terms. $\eta$ drops to $\approx 0.1$ by $w \approx 1.035$ in Fig.~\ref{fig:1}, whereas it only drops to $\approx 0.4$ by $w \approx 1.13$ in Fig.~\ref{fig:4}. This follows directly from the $m_Q^2/\Lambda_B^2$ scaling in the exponential ansatz of Eq.~(\ref{eq:iwf_ansatz}). For the heavier $bbb\rightarrow  bbc$ transition, the accessible $\omega$ window near zero recoil is narrow and $\eta$ already decays steeply within it, whereas for the lighter $bcc\rightarrow  ccc$ transition the accessible $\omega$ range is much larger but $\eta$ declines more slowly. The last column of Table~\ref{tab:Table5} lists the lepton flavour universality
ratio $\mathcal{R}$ for each transition,
\begin{equation}
  \mathcal{R} = \frac{\Gamma_{\Omega_i \rightarrow \Omega_f \tau\bar\nu_\tau}}
                     {\Gamma_{\Omega_i \rightarrow \Omega_f \ell\bar\nu_\ell}},
  \qquad \ell = e,\ \mu,
\end{equation}
where $\Gamma_{\Omega_i \rightarrow \Omega_f e\bar\nu_e}$ and $\Gamma_{\Omega_i \to \Omega_f \mu\bar\nu_\mu}$ are found to be nearly identical across all eight decays, so $\mathcal{R}$ is entirely $\tau$-driven and reflects genuine phase-space and dynamical suppression of the $\tau$ channel relative to the light leptons. $\mathcal{R}$ decreases systematically across the three sectors, from $\Omega_{bbb}^{*}\to\Omega_{bbc}^{(*)}$, through $\Omega_{bbc}^{(*)}\to\Omega_{bcc}^{(*)}$, down to $\Omega_{bcc}^{(*)}\to\Omega_{ccc}^{(*)}$. This follows the same $m_Q$-dependence of the Isgur-Wise function: lighter spectator content widens the accessible $w$-range, and the $\tau$ channel, confined closer to zero recoil by its larger lepton mass, loses a growing share of that range relative to $e$ and $\mu$. Because $\mathcal{R}$ is a ratio of widths (or branching ratios) for a common initial state, the borrowed lifetime $\tau_{\Omega_i}$ used elsewhere in this work cancels out, making $\mathcal{R}$ independent of that external input. The $\mathcal{R}$ values obtained here range from $0.29$ to $0.56$ across the eight transitions studied. No experimental or lattice determination of $\mathcal{R}$ exists for triply heavy baryons. The same type of ratio has been measured for singly heavy baryons, e.g. $\mathcal{R}_{\Lambda_c} = 0.242 \pm 0.026 \pm 0.040 \pm 0.059$ for $\Lambda_b\to\Lambda_c\,\ell\bar\nu_\ell$~\cite{Aaij2022} transition, demonstrating that such ratios are experimentally accessible for $b \rightarrow c$ baryon transitions; the values reported here may serve as a benchmark for future measurements in the triply heavy sector.

The calculated branching ratios are presented in Table \ref{tab:Table6}. $Br_1$ and $Br_2$ in Table \ref{tab:Table6} are the calculated branching ratios using the Leading Order (LO) and Next-to-leading order (NLO) lifetimes quoted in Ref.~\cite{Wang2018}. For $\Omega_{bbc}$ and $\Omega_{bcc}$, we assume $\tau(\Omega^*_{bbc})=\tau(\Omega_{bbc})$ and $\tau(\Omega^*_{bcc})=\tau(\Omega_{bcc})$. This is a reasonable approximation given the mass splitting between the spin partners (Table~\ref{tab:Table2}), although it introduces an additional source of systematic uncertainty beyond the quoted errors. The branching ratios for $\Omega_i \rightarrow \Omega_f e\bar{\nu}_e$ and $\Omega_i \rightarrow \Omega_f \mu \bar{\nu}_{\mu}$ are found to be nearly degenerate in all channels, as expected from the negligible lepton mass suppression at these mass scales, while the branching ratios for $\Omega_i \rightarrow \Omega_f \tau \bar{\nu}_{\tau}$ transitions are suppressed by a factor of 2--3 relative to the light-lepton modes due to phase-space restriction from the tau mass. The largest branching ratios are obtained for the $\Omega_{bcc}\rightarrow\Omega^*_{ccc}$ and $\Omega^*_{bcc}\to\Omega^*_{ccc}$ channels.

\begin{figure*}
\centering
\begin{subfigure}{0.45\textwidth}
\includegraphics[scale=0.3]{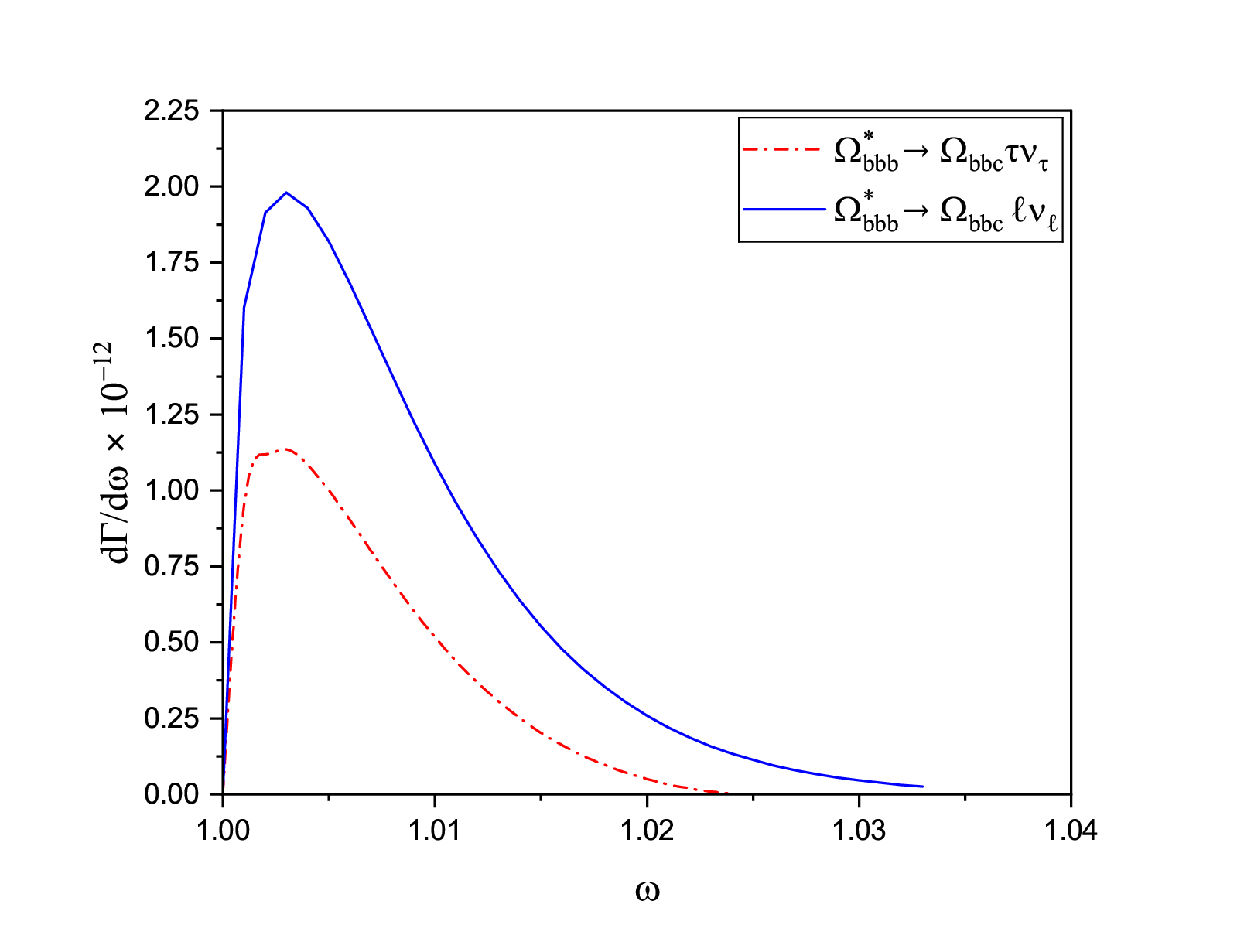}
\caption{$\Omega^*_{bbb} \rightarrow \Omega_{bbc}\,\ell\bar{\nu}$ with ($\ell=e,\mu$)}
\label{fig:5a}
\end{subfigure}
\hfill
\begin{subfigure}{0.45\textwidth}
\includegraphics[scale=0.3]{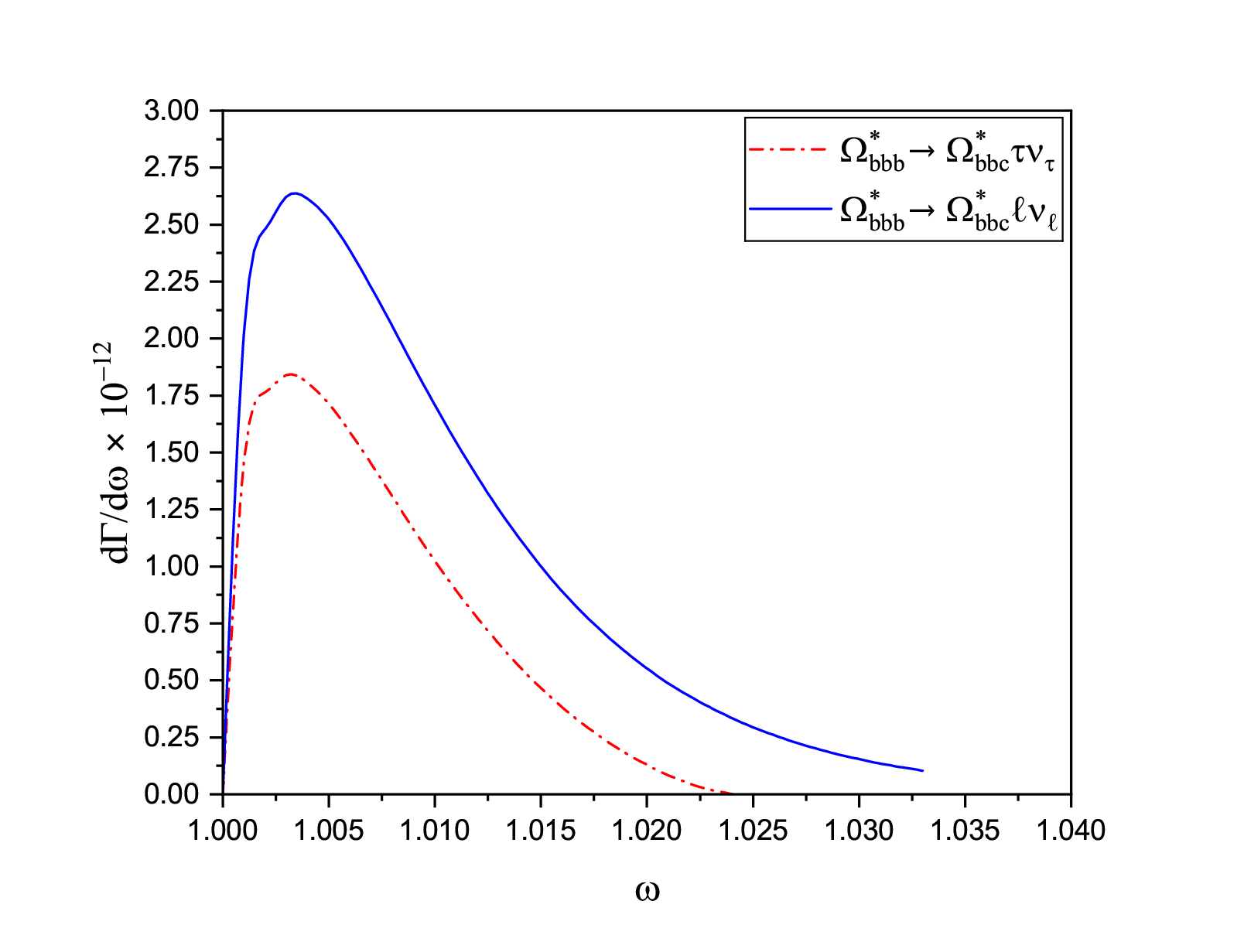}
\caption{$\Omega^*_{bbb} \rightarrow \Omega^*_{bbc}\,\ell\bar{\nu}$ with ($\ell=e,\mu$)}
\label{fig:5b}
\end{subfigure}
\caption{The variation in differential decay width for $\Omega^*_{bbb} \rightarrow \Omega^{(*)}_{bbc}\,\ell\bar{\nu}$ transitions, shown for $\ell\bar{\nu}_\ell$ ($\ell=e,\mu$) and $\tau\bar{\nu}_\tau$ final states}
\label{fig:5}
\end{figure*}

\begin{figure*}
\centering
\begin{subfigure}{0.45\textwidth}
\includegraphics[scale=0.3]{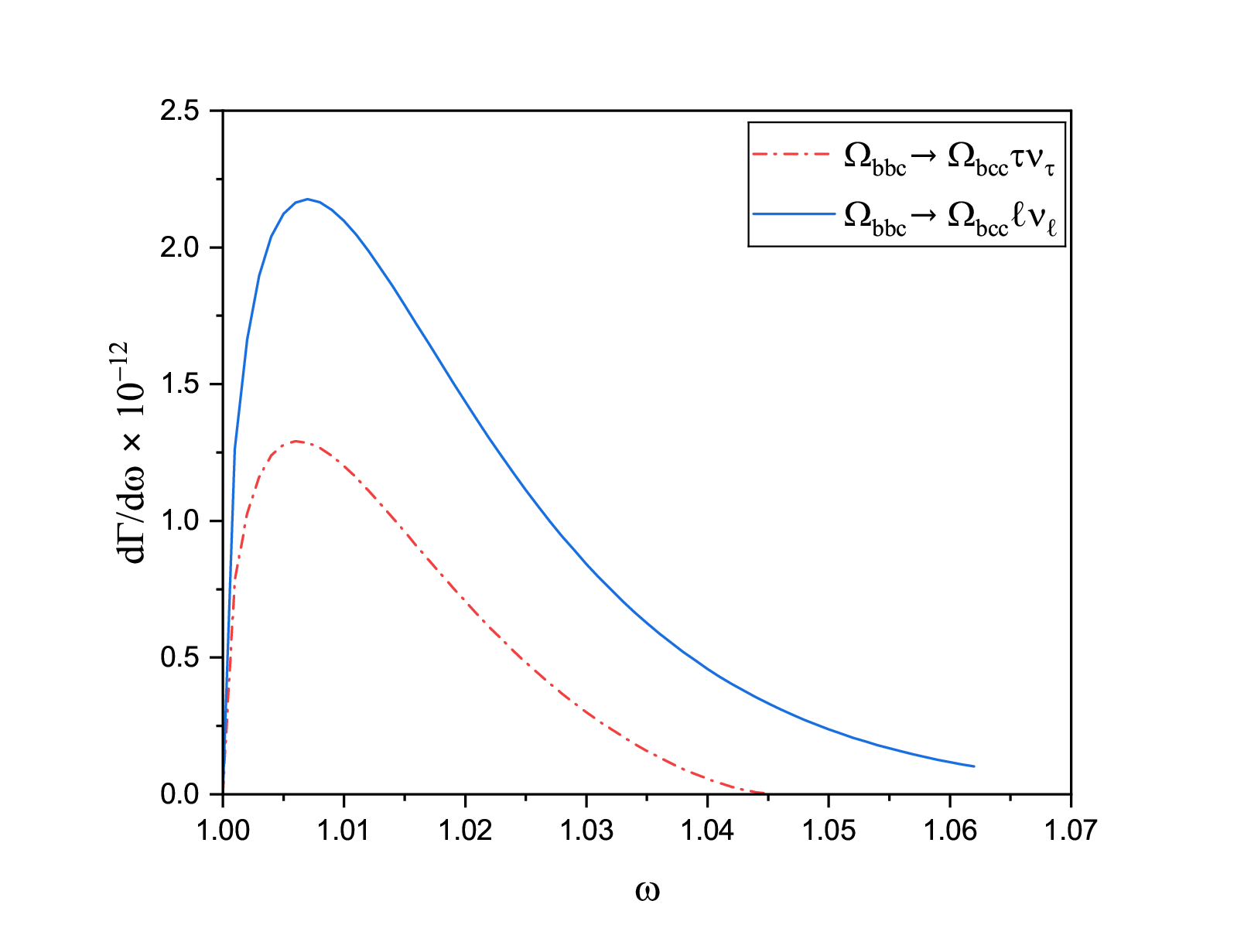}
\caption{$\Omega_{bbc} \rightarrow \Omega_{bcc}\,\ell\bar{\nu}$ with ($\ell=e,\mu$)}
\label{fig:6a}
\end{subfigure}
\hfill
\begin{subfigure}{0.45\textwidth}
\includegraphics[scale=0.3]{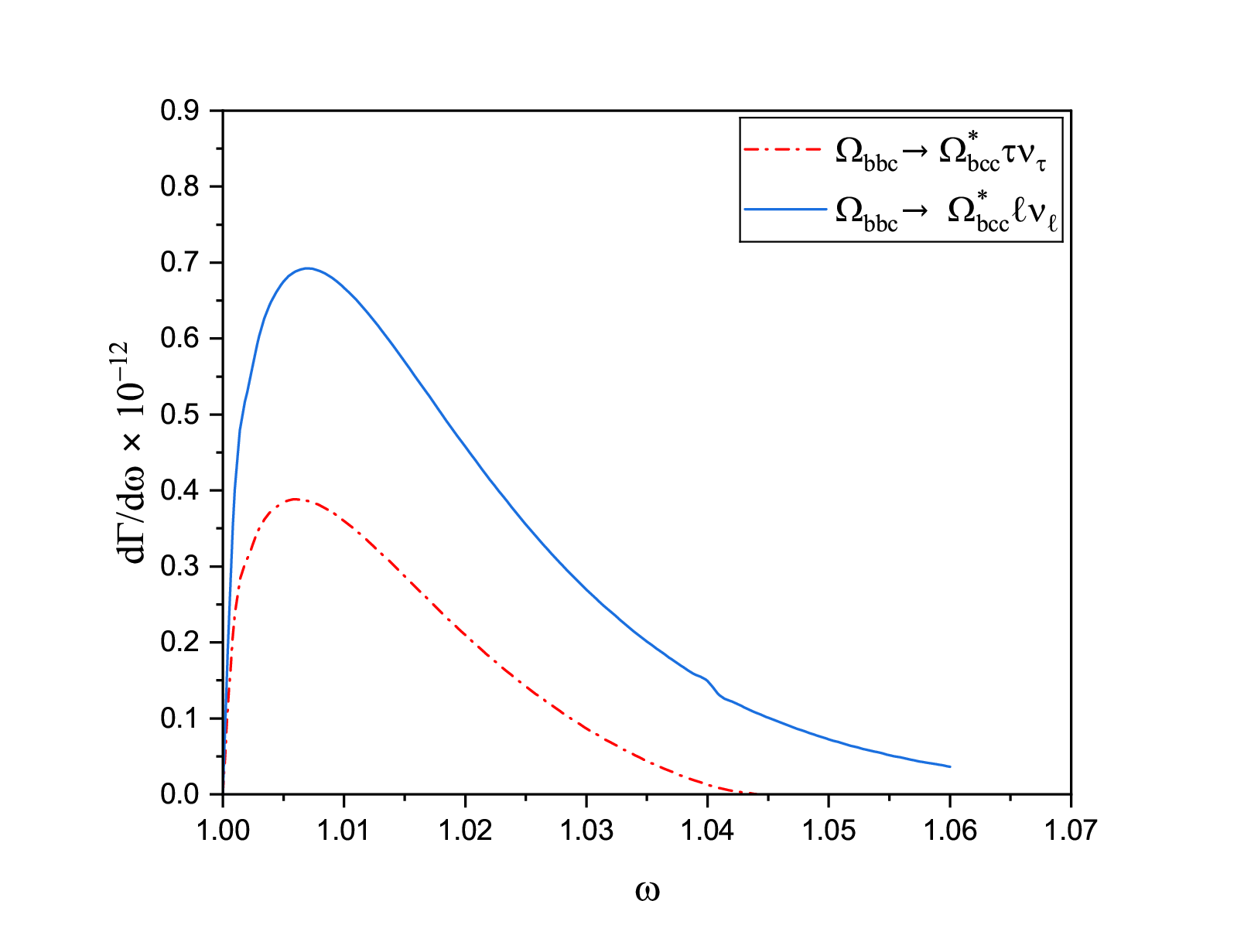}
\caption{$\Omega_{bbc} \rightarrow \Omega^*_{bcc}\,\ell\bar{\nu}$ with ($\ell=e,\mu$)}
\label{fig:6b}
\end{subfigure}
\caption{The variation in differential decay width for $\Omega_{bbc} \rightarrow \Omega^{(*)}_{bcc}\,\ell\bar{\nu}$ transitions, shown for $\ell\bar{\nu}_\ell$ ($\ell=e,\mu$) and $\tau\bar{\nu}_\tau$ final states}
\label{fig:6}
\end{figure*}

\begin{figure*}
\centering
\begin{subfigure}{0.45\textwidth}
\includegraphics[scale=0.3]{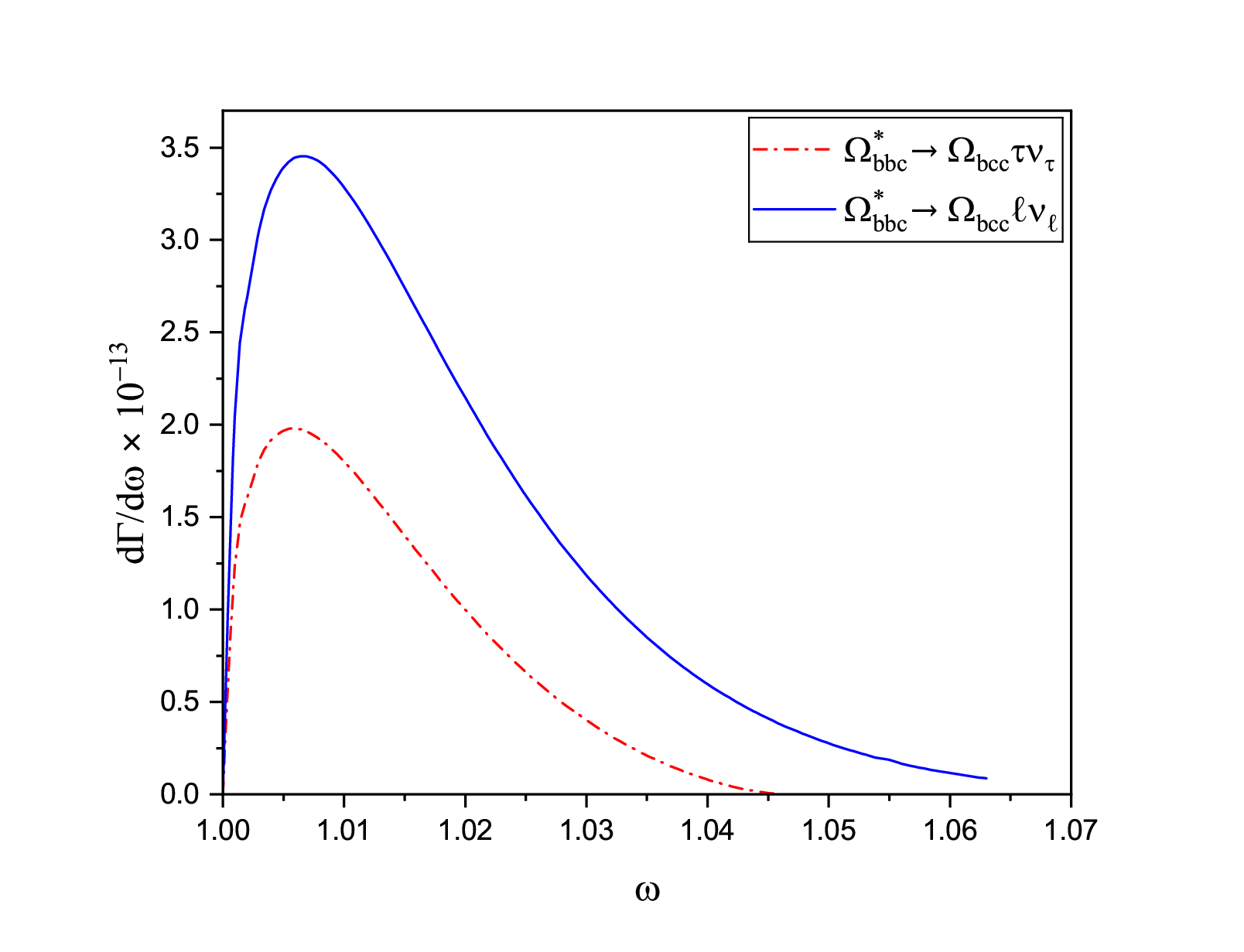}
\caption{$\Omega^*_{bbc} \rightarrow \Omega_{bcc}\,\ell\bar{\nu}$ with ($\ell=e,\mu$)}
\label{fig:7a}
\end{subfigure}
\hfill
\begin{subfigure}{0.45\textwidth}
\includegraphics[scale=0.3]{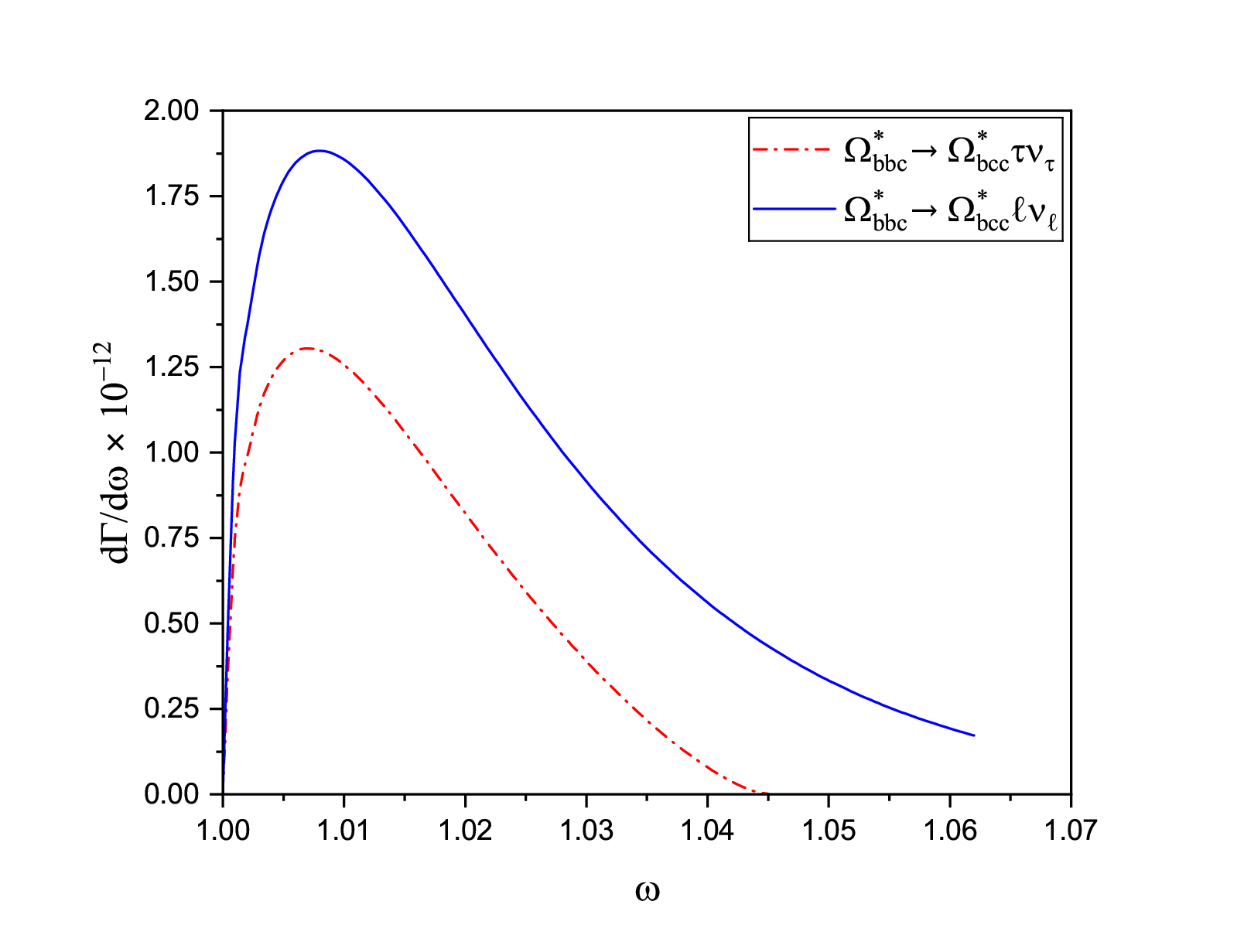}
\caption{$\Omega^*_{bbc} \rightarrow \Omega^*_{bcc}\,\ell\bar{\nu}$ with ($\ell=e,\mu$)}
\label{fig:7b}
\end{subfigure}
\caption{The variation in differential decay width for $\Omega^*_{bbc} \rightarrow \Omega^{(*)}_{bcc}\,\ell\bar{\nu}$ transitions, shown for $\ell\bar{\nu}_\ell$ ($\ell=e,\mu$) and $\tau\bar{\nu}_\tau$ final states}
\label{fig:7}
\end{figure*}

\begin{figure*}
\centering
\begin{subfigure}{0.45\textwidth}
\includegraphics[scale=0.3]{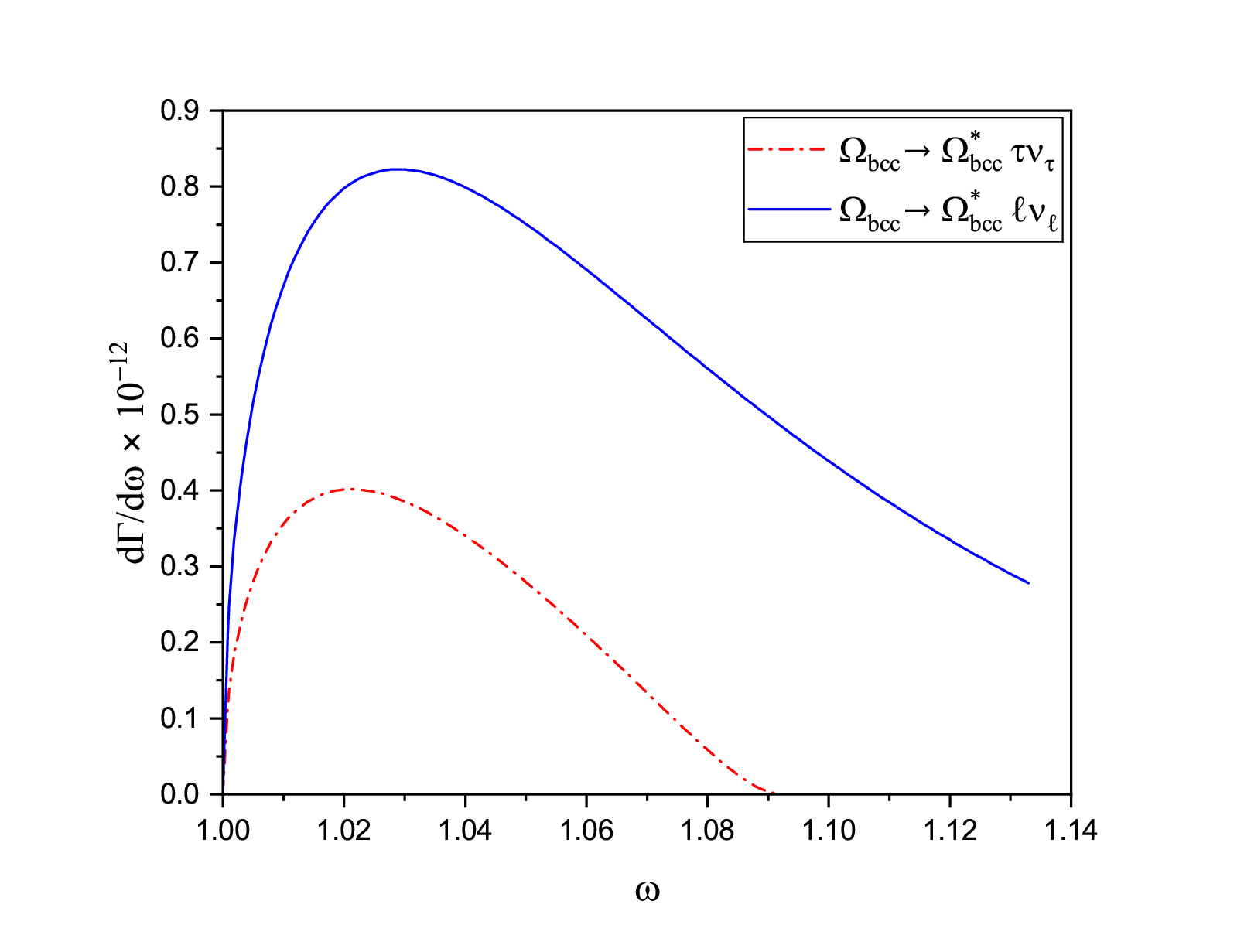}
\caption{$\Omega_{bcc} \rightarrow \Omega^*_{ccc}\,\ell\bar{\nu}$}
\label{fig:8a}
\end{subfigure}
\hfill
\begin{subfigure}{0.45\textwidth}
\includegraphics[scale=0.3]{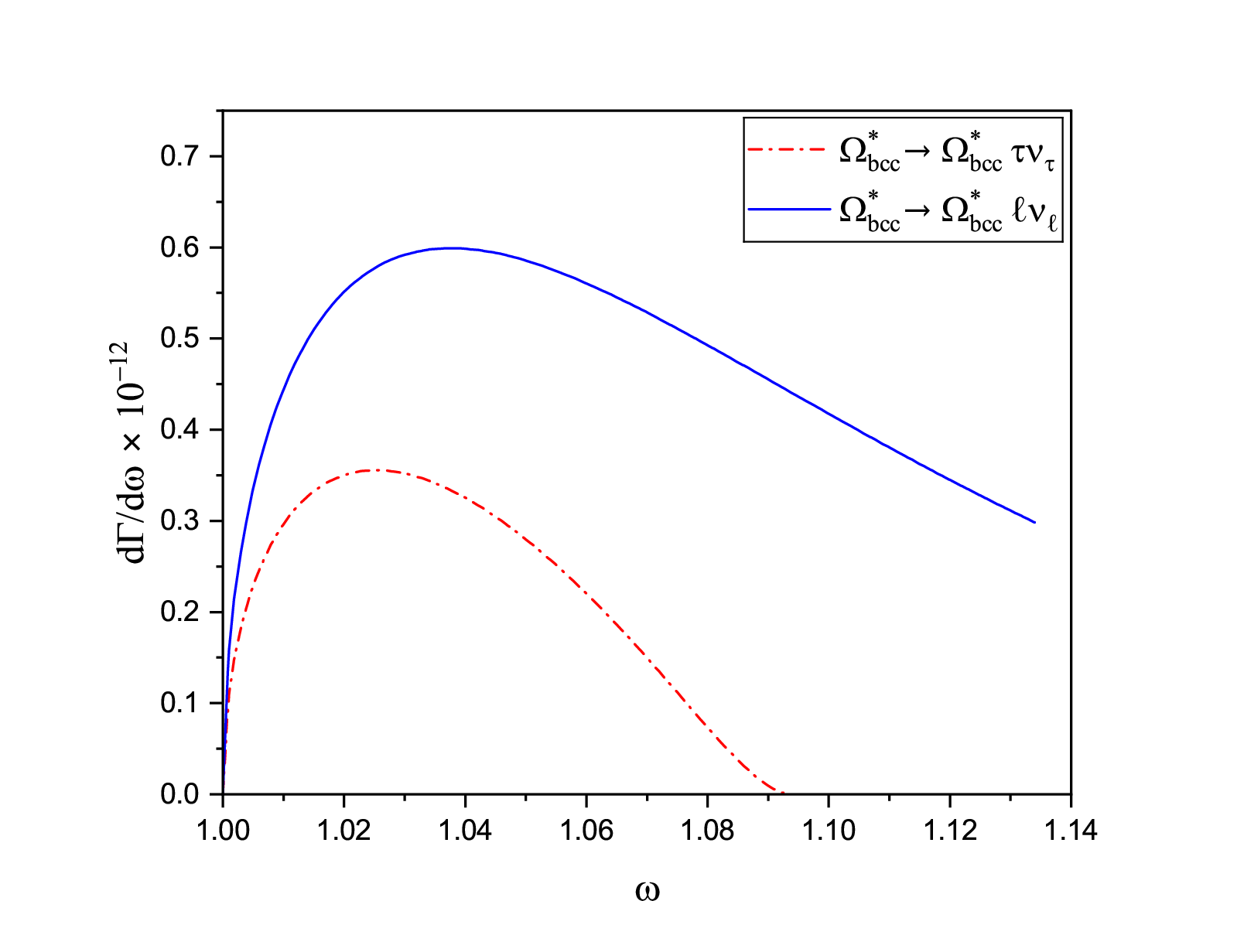}
\caption{$\Omega^*_{bcc} \rightarrow \Omega^*_{ccc}\,\ell\bar{\nu}$}
\label{fig:8b}
\end{subfigure}
\caption{The variation in differential decay width for $\Omega^{(*)}_{bcc} \rightarrow \Omega^*_{ccc}\,\ell\bar{\nu}$ transitions, shown for $\ell\bar{\nu}_\ell$ ($\ell=e,\mu$) and $\tau\bar{\nu}_\tau$ final states}
\label{fig:8}
\end{figure*}

\begin{table*}[h]
\caption{\label{tab:Table2} Ground-state masses of triply heavy baryons (in GeV).}
\begin{tabular}{lcccccc}
\hline\noalign{\smallskip}
Reference & $\Omega^*_{bbb}$ & $\Omega^*_{bbc}$ & $\Omega_{bbc}$ & $\Omega^*_{bcc}$ & $\Omega_{bcc}$ & $\Omega^*_{ccc}$ \\
\hline\noalign{\smallskip}
Our & 14.852 & 11.461 & 11.430 & 8.059 & 8.031 & 4.825 \\
LQCD \cite{Brown2014} & 14.366 & 11.229 & 11.195 & 8.037 & 8.007 & 4.796 \\
Variational \cite{Flynn2012} & 14.398 & 11.245 & 11.214 & 8.046 & 8.018 & 4.799 \\
RQM \cite{Faustov2022} & 14.468 & 11.217 & 11.198 & 7.999 & 7.984 & 4.712 \\
\hline\noalign{\smallskip}
\end{tabular}
\end{table*}

\begin{table*}[h]
\caption{\label{tab:Table3a} Magnetic moments of triply heavy baryons (in units of nuclear magneton $\mu_N$).}
\tiny
\begin{tabular}{lcccccccc}
\hline\noalign{\smallskip}
Baryon & Our & \cite{Simonis2018} & PM(BD) \cite{Brac1996} & PM(AL1) \cite{Brac1996}& PM \cite{Thakkar2016}& Hyp \cite{Patel2009} & EM\&C \cite{Dhir2009,Dhir2013} & LC-QCDSR\cite{Mutuk2022} \\
\hline\noalign{\smallskip}
$\Omega^{*}_{bbb}$   & $-0.189$  & $-0.178$ & $-0.178$ & $-0.180$ & $-0.196$ & $-0.195$ & $-0.198$ & $-$ \\
$\Omega_{bbc}$       & $-0.219$ & $-0.187$ & $-0.191$ & $-0.193$ & $-0.223$ & $-0.203$ & $-0.200$ & $-0.19\pm0.06$ \\
$\Omega^{*}_{bbc}$   & 0.279  & 0.204    & $-$ & $-$ & 0.285    & 0.216    & 0.225    & $-$ \\
$\Omega_{bcc}$       & 0.560     & 0.455    & 0.466    & 0.475    & 0.565    & 0.502    & 0.522    & $0.61\pm0.21$ \\
$\Omega^{*}_{bcc}$   & 0.743     & 0.594    & $-$ & $-$ & 0.751    & 0.651    & 0.703    & $-$ \\
$\Omega^{*}_{ccc}$   & 1.166     & 0.989    & 1.00     & 1.02     & 1.18     & 1.19     & 1.16     & $-$ \\
\hline\noalign{\smallskip}
\end{tabular}
\end{table*}

\begin{table*}[h]
\centering
\caption{\label{tab:Table4} Transition magnetic moments (in nuclear magneton $\mu_N$) and radiative $M1$ decay widths (in keV) of triply heavy baryons.}
\begin{tabular}{lcccc}
\hline\noalign{\smallskip}
 & \multicolumn{2}{c}{$\mu\,(B^*\rightarrow B)$} & \multicolumn{2}{c}{$\Gamma\,(B^*\rightarrow B\gamma)$} \\
\cmidrule(lr){2-3} \cmidrule(lr){4-5}
 & $\Omega^{*}_{bbc}\rightarrow\Omega_{bbc}\,\gamma$ & $\Omega^{*}_{bcc}\rightarrow\Omega_{bcc}\,\gamma$ & $\Omega^{*}_{bbc}\rightarrow\Omega_{bbc}\,\gamma$ & $\Omega^{*}_{bcc}\rightarrow\Omega_{bcc}\,\gamma$ \\
\hline\noalign{\smallskip}
Our & $-0.439$ & $0.440$ & $0.007$ & $0.0195$ \\
\cite{Hazra2021} & $-0.41709\pm0.0005$ & $0.41709\pm0.0008$ & $0.0282\pm0.0008$ & $0.0192\pm0.0006$ \\
\cite{Simonis2018} & $-0.352$ & $0.362$ & $0.013$ & $0.010$ \\
\hline\noalign{\smallskip}
\end{tabular}
\end{table*}

\begin{table}[h]
    \caption{\label{tab:Table5} Semileptonic decay widths of triply heavy baryons (in $10^{10}~\mathrm{s}^{-1}$) and Lepton flavour universality ratio $R$.}
    \begin{tabular}{ccccc}
    \hline\noalign{\smallskip}
    Decay & $\Gamma_{e,\mu}$ & $\Gamma_\tau$ & $\Gamma_e$ \cite{Flynn2012} & $\mathcal{R}$ \\
    \hline\noalign{\smallskip}
    $\Omega_{bbb}^{*}\rightarrow\Omega_{bbc}$        & 3.54  & 1.70  & 3.95 & 0.48 \\
    $\Omega_{bbb}^{*}\rightarrow\Omega_{bbc}^{*}$    & 5.50  & 3.07  & 6.34 & 0.56 \\
    \hline\noalign{\smallskip}
    $\Omega_{bbc}\rightarrow\Omega_{bcc}$            & 9.07  & 4.19  & 7.98 & 0.46 \\
    $\Omega_{bbc}\rightarrow\Omega_{bcc}^{*}$        & 2.89  & 1.24  & 2.42 & 0.43 \\
    $\Omega_{bbc}^{*}\rightarrow\Omega_{bcc}$        & 1.35  & 0.62  & 1.17 & 0.46 \\
    $\Omega_{bbc}^{*}\rightarrow\Omega_{bcc}^{*}$    & 8.94  & 4.59  & 7.74 & 0.51 \\
    \hline\noalign{\smallskip}
    $\Omega_{bcc}\rightarrow\Omega_{ccc}^{*}$        & 11.8  & 3.43  & 8.01 & 0.29 \\
    $\Omega_{bcc}^{*}\rightarrow\Omega_{ccc}^{*}$    & 9.64  & 3.23  & 6.28 & 0.34 \\
    \hline\noalign{\smallskip}
    \end{tabular}
\end{table}

\begin{table*}[h]
    \caption{\label{tab:Table6} Branching ratios (\%) of triply heavy baryons for the $b\rightarrow c$ semileptonic decay channel, using leading-order (LO) and next-to-leading-order (NLO) lifetimes from Ref.~\cite{Wang2018}. Uncertainties are propagated from the quoted lifetime uncertainties only. Branching ratios to $e$ and $\mu$ final states coincide in this approximation and are quoted jointly as $\mathrm{Br}_{e}$.}
\begin{tabular}{ccccc}
\hline\noalign{\smallskip}
 & \multicolumn{2}{c}{$\mathrm{Br}_{e,\mu}$ (\%)} & \multicolumn{2}{c}{$\mathrm{Br}_{\tau}$ (\%)} \\
Decay & $\mathrm{Br}_1$ & $\mathrm{Br}_2$ & $\mathrm{Br}_1$ & $\mathrm{Br}_2$ \\
\hline\noalign{\smallskip}
$\Omega^*_{bbb}\rightarrow\Omega_{bbc}$        & $1.59\pm0.11$ & $1.20\pm0.14$ & $0.77\pm0.05$ & $0.58\pm0.07$ \\
$\Omega^*_{bbb}\rightarrow\Omega^*_{bbc}$      & $2.48\pm0.17$ & $1.87\pm0.22$ & $1.38\pm0.09$ & $1.04\pm0.12$ \\
\hline\noalign{\smallskip}
$\Omega_{bbc}\rightarrow\Omega_{bcc}$          & $3.48\pm0.30$ & $2.28\pm0.27$ & $1.61\pm0.14$ & $1.05\pm0.13$ \\
$\Omega_{bbc}\rightarrow\Omega^*_{bcc}$        & $1.11\pm0.10$ & $0.72\pm0.09$ & $0.48\pm0.04$ & $0.31\pm0.04$ \\
$\Omega^*_{bbc}\rightarrow\Omega_{bcc}$        & $0.52\pm0.04$ & $0.34\pm0.04$ & $0.24\pm0.02$ & $0.16\pm0.02$ \\
$\Omega^*_{bbc}\rightarrow\Omega^*_{bcc}$      & $3.43\pm0.30$ & $2.24\pm0.27$ & $1.76\pm0.15$ & $1.15\pm0.14$ \\
\hline\noalign{\smallskip}
$\Omega_{bcc}\rightarrow\Omega^*_{ccc}$        & $3.99\pm0.41$ & $2.34\pm0.30$ & $1.16\pm0.12$ & $0.68\pm0.09$ \\
$\Omega^*_{bcc}\rightarrow\Omega^*_{ccc}$      & $3.26\pm0.34$ & $1.91\pm0.24$ & $1.09\pm0.11$ & $0.64\pm0.08$ \\
\hline\noalign{\smallskip}
\end{tabular}
\end{table*}

\section{Conclusions}
We presented a comprehensive study of the static and dynamic properties of triply heavy baryons within the hypercentral constituent quark model, computing their ground-state masses, magnetic moments, transition magnetic moments, radiative $M1$ decay widths, and exclusive semileptonic $b\rightarrow  c$ decay widths. The calculated ground-state masses of all triply heavy baryons agree reasonably well with the predictions. The transition magnetic moments and radiative $M1$ widths are consistent with other model predictions, with the remaining discrepancies in the radiative widths attributable to their cubic dependence on the photon momentum, which amplifies small differences in the predicted mass splitting between the initial and final states. For semileptonic decays, the hadronic matrix elements of the $b\rightarrow  c$ transitions are reduced to a single Isgur--Wise function using heavy-quark spin symmetry. The lepton flavour universality ratio $\mathcal{R}$, decreasing systematically from $0.56$ to $0.29$ across the three sectors, is predicted here for the first time and may serve as a benchmark for future measurements.

\end{sloppypar}
\end{document}